\documentclass[twocolumn,english,aps,pra,longbibliography, superscriptaddress]{revtex4-1}
\usepackage[latin9]{inputenc}
\usepackage{amsmath}
\usepackage{amssymb}
\usepackage{braket}
\usepackage{color}
\usepackage{graphicx}
\usepackage{pifont}
\usepackage{xcolor}
\usepackage[colorlinks=true,urlcolor=blue,citecolor=blue,linkcolor=blue]{hyperref}
\usepackage{setspace}
\usepackage{siunitx} 

\begin{document}

\title{Minimum quantum run-time characterization and calibration via restless measurements with dynamic repetition rates}

\author{Caroline Tornow}
\affiliation{ETH Zurich 8093, Switzerland}
\affiliation{IBM Quantum -- IBM Research - Zurich, S\"aumerstrasse 4, 8803 R\"uschlikon, Switzerland}
\author{Naoki Kanazawa}
\affiliation{IBM Quantum -- IBM Research Tokyo, Tokyo, 103-8510, Japan}
\author{William E. Shanks}
\affiliation{IBM Quantum -- IBM T.J. Watson Research Center, Yorktown Heights, New York 10598, USA}
\author{Daniel J. Egger}
\email{deg@zurich.ibm.com}
\affiliation{IBM Quantum -- IBM Research - Zurich, S\"aumerstrasse 4, 8803 R\"uschlikon, Switzerland}
\date{\today}

\begin{abstract}
    The performance of a quantum processor depends on the characteristics of the device and the quality of the control pulses.
    Characterizing cloud-based quantum computers and calibrating the pulses that control them is necessary for high-fidelity operations.
    However, this time intensive task eats into the availability of the device.
    Here, we show restless measurements with a dynamic repetition rate that speed-up calibration and characterization tasks.
    Randomized benchmarking is performed 5.3 times faster on the quantum device than when an active reset is used and without discarding any data.
    In addition, we calibrate a qubit with parameter scans and error-amplifying gate sequences and show speed-ups of up to a factor of forty on the quantum device over active reset.
    Finally, we present a methodology to perform restless quantum process tomography that mitigates restless state preparation errors.
    These results reduce the footprint of characterization and calibration tasks.
    Quantum computers can thus either spend more time running applications or run calibrations more often to maintain gate fidelity.
\end{abstract}

\maketitle

\section{Introduction}

Quantum computers have the potential to impact a wide range of applications~\cite{Farhi2014, Biamonte2017, Moll2018, Kandala2018, Braine2019, Egger2020}.
Their performance is measurable along three dimensions, quality, scale and speed~\cite{Wack2021}.
Quantum Volume~\cite{Cross2019, Jurcevic2021}, for example, measures scale and quality.
However, maintaining the quality of a quantum computer requires periodic gate calibration and characterization performed by jobs interleaved with normal circuit execution.
These jobs therefore take up time on the quantum processor. 
Reducing the time they take increases system availability and enables more frequent calibration, thus resulting in higher quality.

High quality gates require carefully designed pulse shapes, such as DRAG~\cite{Motzoi2009}, calibrated to the quantum system.
Analytic pulses with few parameters are typically calibrated using gate sequences tailored to amplify different errors such as over- or under-rotations and phase errors~\cite{Sheldon2016}.
Quantum optimal control~\cite{Glaser2015} provides methods to design gates~\cite{Schutjens2013, Heeres2017, Abdelhafez2020}, reduce pulse duration~\cite{Egger2013, Kirchhoff2018, Werninghaus2020} and increase fidelity~\cite{Kelly2014}.
However, to overcome drifts and model inaccuracies these methods often require data-intensive closed-loop optimization~\cite{Egger2014, Kelly2014, Werninghaus2020}.
This makes fewer control parameters~\cite{Machnes2018} and high-speed data acquisition desirable~\cite{Rol2017, Werninghaus2021}.
Optimal control requires a fidelity metric to optimize.
Randomized benchmarking (RB) measures the average gate fidelity~\cite{Magesan2011, Magesan2012b, Corcoles2013} and provides a state preparation and measurement (SPAM) error insensitive metric to both optimize quantum gates~\cite{Kelly2014, Rol2017, Werninghaus2020} and learn the underlying system model~\cite{Wittler2021}.
By contrast, quantum process tomography (QPT) measures the gate fidelity~\cite{OBrien2004, Mohseni2008, Bialczak2010, Nielsen2010, Pechal2020, Earnest2021, kiktenko2021} and provides more information on the process matrix but is sensitive to SPAM errors.
Combining RB with quantum process tomography improves model learning~\cite{Wittler2021}.

\begin{figure}[htbp!]
    \centering
    \includegraphics[width=1\columnwidth, clip, trim= 0 5 0 0]{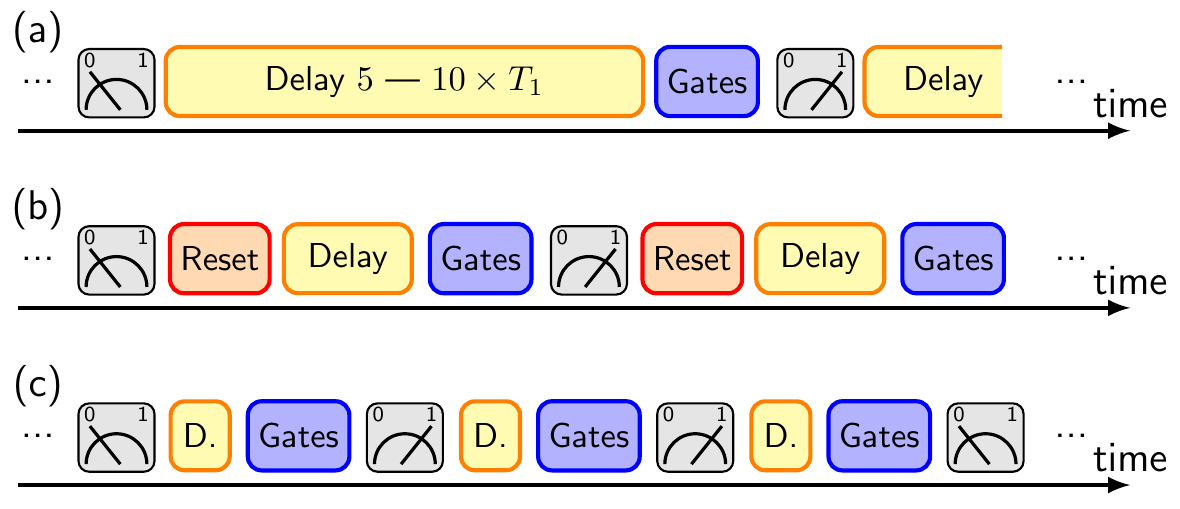}
    \caption{
    Illustration of qubit initialization schemes.
    Time axes are not to scale.
    (a) The qubit is passively reset to the ground state by waiting five to ten times the $T_1$-time. 
    (b) The qubit is actively reset after each measurement.
    A $50$ to $250\,\mu{\rm s}$ delay is often necessary to ensure a high reset quality.
    (c) Restless measurements where the outcome of the previous measurement initializes the qubit to either $\ket{0}$ or $\ket{1}$.
    A short delay of typically $1\,\mu{\rm s}$ is sometimes needed to allow time for classical processing of the readout signal.
    }
    \label{fig:init}
\end{figure}

Speed is influenced by both classical and quantum elements~\cite{Wack2021} such as run-time compilation and the ability to quickly reset the qubits.
Superconducting qubit-based quantum processors~\cite{Devoret2013, Krantz2019, Blais2021} enjoy long coherence times~\cite{Rigetti2012, Place2021} and comparatively short gates.
Long $T_1$ times make active reset necessary since waiting $5$ to $10$ times $T_1$ is inefficient~\cite{Wack2021}, see Fig.~\ref{fig:init}(a).
However, active reset also requires time and typically lasts a few microseconds~\cite{Riste2012, Govia2015, Geerlings2013, Magnard2018, Egger2018, Corcoles2021}.
For example, the qubits can be unconditionally reset by applying a $\pi$-pulse to the $\ket{1}$ to $\ket{2}$ transiton of the transmon and then driving the transition between the $\ket{2}$ state of the transmon and the first excited state of the readout resonator where the excitation quickly decays to the environment~\cite{Magnard2018, Egger2018}.
Furthermore, a delay is often necessary after a reset operation to avoid deteriorating the initialization quality~\cite{Egger2018}, see Fig.~\ref{fig:init}(b).
For example, on IBM Quantum systems this configurable delay currently has a default value of $50$ to $250\,{\rm \mu s}$ depending on the system~\cite{Wack2021}.
In restless measurements the qubits are not reset in between circuits. 
This provides a fast alternative to several characterization and calibration tasks~\cite{Rol2017, Werninghaus2020, Werninghaus2021}.
Here, the outcome of the projective measurement at the end of a circuit initializes the qubits to either $\ket{0}$ or $\ket{1}$ for the next circuit, see Fig.~\ref{fig:init}(c).
For a large class of circuits the states 0 and 1 can be relabeled without modifying the circuit.

Quantum circuits can either be executed with a static or a dynamic repetition rate.
With a static repetition rate $R$ each circuit must fit within the period $1/R$. 
By contrast, when the repetition rate is dynamic a quantum circuit will begin executing a fixed delay after the previous circuit completed.
Here, we demonstrate that a dynamic repetition rate improves the quality of restless measurements.
Indeed, restless measurements with a static repetition rate $R$ must fit all gates and readout operations for each circuit within a period $1/R$~\cite{Werninghaus2021}.
This produces a variable delay after each measurement when the circuits have a different duration as in RB and therefore a variable state preparation error.
By contrast, when the repetition rate is dynamic there is a short fixed delay after each circuit and consequently a fixed state preparation error.
In Sec.~\ref{sec:restless} we review restless measurements.
Next, in Sec.~\ref{sec:rb}, we show that by using dynamic repetition rates all of the restless randomized benchmarking data is usable, as opposed to discarding $60\%$ of it as in Ref.~\cite{Werninghaus2021}.
In Sec.~\ref{sec:cal}, we show restless qubit calibration with error amplifying gate sequences.
Furthermore, we show in Sec.~\ref{sec:qpt} how to perform QPT with restless measurements and how to mitigate state preparation errors.
We conclude in Sec.~\ref{sec:conclusion}.

\section{Restless measurements\label{sec:restless}}

IBM Quantum systems are built with fixed-frequency transmon qubits dispersively coupled to readout resonators.
The $N$ measurement shots of $K$ circuits are gathered by sequentially measuring each circuit and repeating this $N$ times.
By default, these systems reset the qubits to their ground state and introduce a fixed delay after each measurement.
This reset can be turned off and the delay reduced from $250\,\mu{\rm s}$ to a $\mu{\rm s}$ or less to perform restless measurements with a dynamic repetition rate, see details in Appendix~\ref{sec:restless_delay}.
Now, if a qubit is measured in state $\ket{i}$ for circuit $k-1$ with $i\in\{0,1\}$ then the initial state of the next circuit $k$ is also $\ket{i}$.
Therefore, the measured outcomes of restless experiments require post-processing.

\subsection{Restless data post-processing\label{sec:data_processing}}

Each qubit is measured by digitizing and then integrating a measurement pulse transmitted through the readout resonator.
Each shot therefore results in a single complex number represented as an in-phase and quadrature component in the IQ plane. These IQ points are discriminated into $\ket{0}$ or $\ket{1}$ states when a trained discriminator is available~\cite{Alexander2020}.
By default, running $K$ circuits with $N$ shots and $n$ qubits on an IBM Quantum system returns $K$ count dictionaries $\{i: C_{i, k} \}_k$. 
Here, the count $C_{i, k}$ with $i\in\{0,1\}^n$ is the number of times state $\ket{i}$ was measured for circuit $k=1,...,K$. 
Furthermore, the counts sum to the total number of shots, e.g. $C_{00, k} + C_{01, k}+C_{10, k} + C_{11, k} = N$ when $n=2$.

\begin{figure}
    \centering
    \includegraphics[width=\columnwidth]{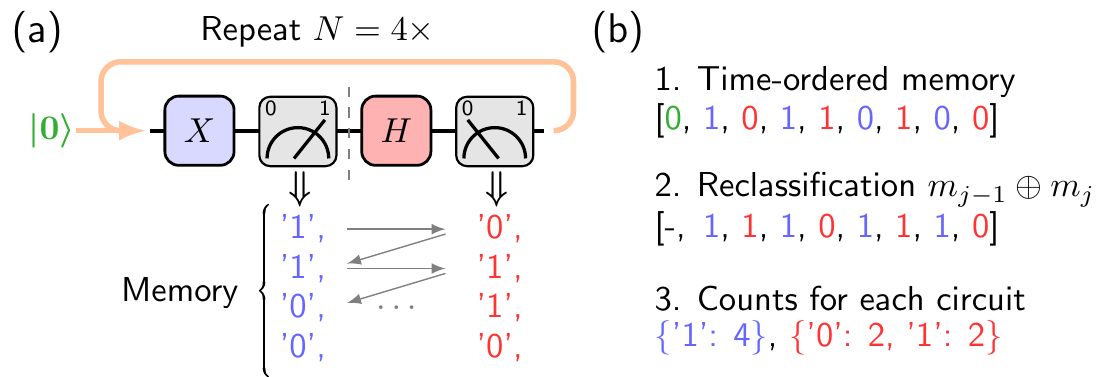}
    \caption{
    Illustration of the restless data post-processing of a single-qubit job with two circuits and four shots.
    (a) The qubit begins in a known state, typically $\ket{0}$.
    The first circuit is an $X$-gate followed by a measurement and the second circuit is a Hadamard gate followed by a measurement.
    The delays are not shown.
    The measurement outcomes are color coded according to the circuit that generated them.
    (b) The measurement outcomes are time-ordered, reclassified with XOR, and summed into count dictionaries.
    }
    \label{fig:data_processing}
\end{figure}

These count dictionaries are unusable in restless measurements.
Instead, to post-process restless data we require the measurement outcome of each shot which the backends can return as a list of outcomes grouped by circuit called the \emph{memory}, see Fig.~\ref{fig:data_processing}.
The state before the very first circuit is typically $\ket{0}$.
The execution begins and each restless single-qubit measurement generates a bit which is appended to the memory of its corresponding circuit, see Fig.~\ref{fig:data_processing}(a).
The memory is post-processed in three steps, see Fig.~\ref{fig:data_processing}(b).
First, the bits are sorted according to the order in which they were acquired.
We label these time-ordered bits $m_j$ with $m_{-1}=0$ to represent the ground state before the first circuit.
Next, we reclassify each bit with the exclusive OR operation to identify state changes, i.e. $m'_j=m_{j-1}\oplus m_j$.
Measurements for which the state changed are identified as the 1 state and those with no change as the 0 state.
These new classified states are then analyzed as normal by building count dictionaries.
This process can be generalized to certain multi-qubit cases by analyzing the outcome of each qubit independently of the others.

Previous work required additional post-processing to account for qubit decay~\cite{Werninghaus2021}.
However, in this work, because of improved qubit lifetimes and reduced delay times, the error due to qubit decay is negligible compared to readout errors and so no additional correction is needed.
Furthermore, such errors are typically absorbed by fit parameters in RB and calibration experiments.

\subsection{Speed-up\label{sec:speed_up}}

We compare each restless experiment to a standard one with active qubit reset and interleave standard and restless jobs to avoid biases due to eventual drifts.
In particular, we compare the time spent by the quantum processor executing the quantum circuits in both settings.
This time is given by
\begin{align}
\label{eq:speed_up}
    \tau^{(x)}=N K\left(\tau^{(x)}_\text{reset}+\tau^{(x)}_\text{delay}+\langle{\tau}_\text{circ}\rangle+\tau_\text{meas}\right),
\end{align}
where $\tau^{(x)}_\text{reset}$ and $\tau^{(x)}_\text{delay}$ are the reset and delay times respectively.
Here, the measurement time $\tau_\text{meas}$ includes the measurement pulse and a short delay of typically $10/(2\pi\kappa)$ that allows the resonator to depopulate where $\kappa$ is the resonator linewidth.
The superscript $(x)$ indicates restless $(r)$ or standard $(s)$ measurements.
The average duration of all $K$ circuits in an experiment is $\langle{\tau}_\text{circ}\rangle=K^{-1}\sum_{k=1}^{K} \tau_{\text{circ},k}$ where $\tau_{\text{circ},k}$ is the duration of only the gates in circuit $k$.
We therefore compute the quantum processor speed-up of restless measurements as $\tau^{(\text{s})}/\tau^{(\text{r})}$ which is independent of the number of circuits and shots due to Eq.~(\ref{eq:speed_up}).
In the terminology of Ref.~\cite{Wack2021}, this speed-up considers the circuit delay and circuit execution times but not the run-time compilation and data transfer times.

\section{Randomized benchmarking\label{sec:rb}}

\begin{figure}[htbp!]
    \centering
    \includegraphics[width=1\columnwidth]{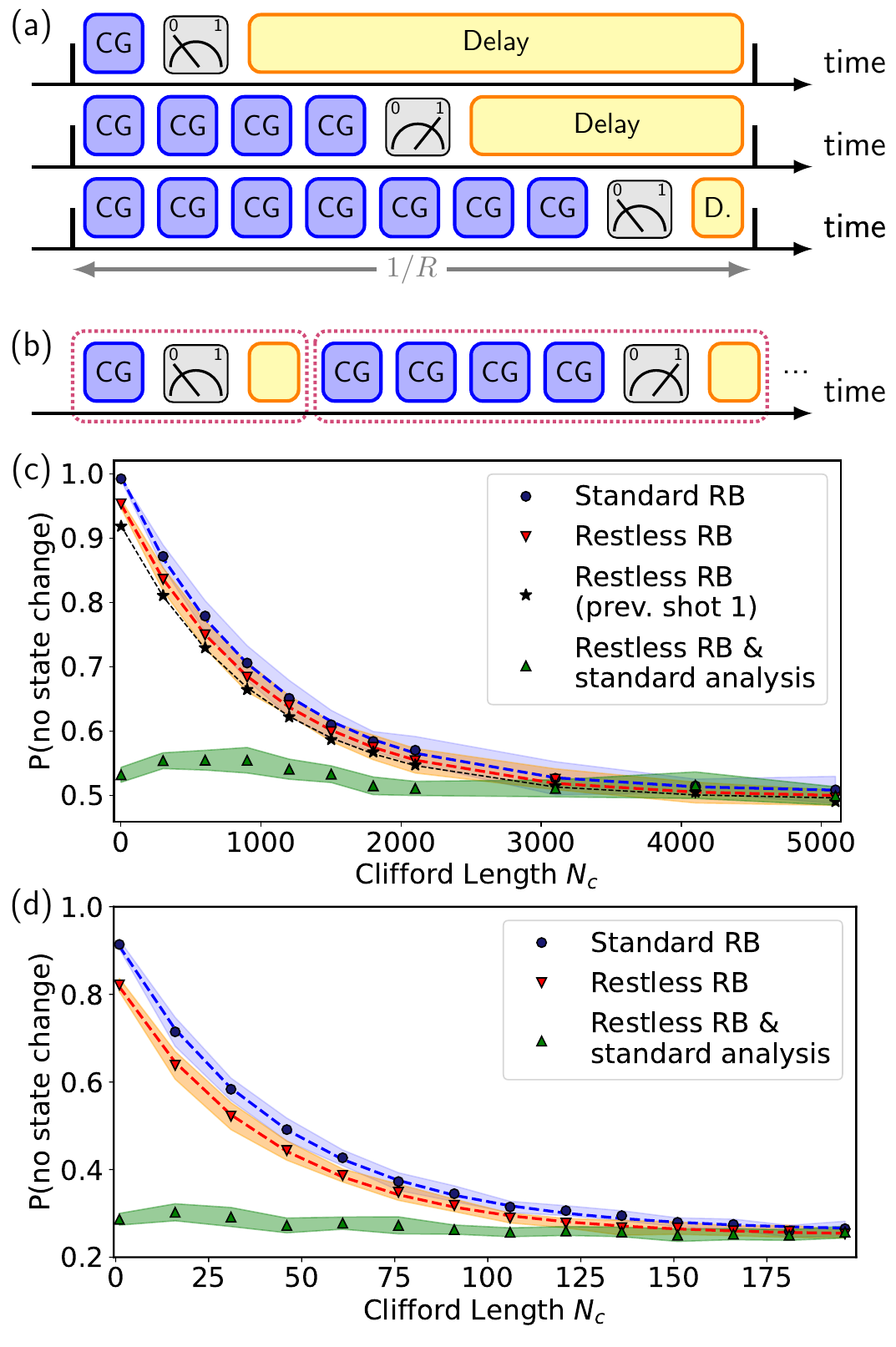}
    \caption{Standard and restless RB.
    (a) With a fixed rate (thick black ticks), there is a variable delay which depends on the number of Clifford gates (blue squares).
    (b) With a dynamic repetition rate the  delay after each measurement is identical for each circuit.
    (c) Single-qubit and (d) two-qubit standard (blue dots) and restless RB (red triangles) with dynamic repetition rates.
    The green triangles show restless data analyzed with the standard data processor.
    The markers are the mean values of ten random realizations of Clifford sequences and the shaded areas are the standard deviation.
    The dotted line with star markers in (c) shows restless RB post-selected to include only the data for which the qubit begins in state $\ket{1}$.
    }
    \label{fig:restless_rb}
\end{figure}

In standard RB the probability of the ground state $P_0$ is measured after a sequence of $N_c$ Clifford gates that compose to the identity.
Fitting $P_0$ to $A \alpha^{N_c} + B$, where $A$, $B$ and $\alpha$ are fit parameters, yields the average error per Clifford gate (EPC) as $(1-\alpha)/2$ and $3(1-\alpha)/4$ for single- and two-qubit RB, respectively.
Restless RB measurements with a fixed trigger rate $R$ only reproduce the average gate fidelity if the outcomes where a qubit began in the excited state are discarded~\cite{Werninghaus2021}.
This is because the variable length of the Clifford sequences creates a variable delay when $1/R$ is constant.
The consequence is a state preparation error dependent on the number of Clifford gates, see Fig.~\ref{fig:restless_rb}(a).
Therefore, 60\% of the restless data in Ref.~\cite{Werninghaus2021} was discarded lowering the effective restless repetition rate by a factor of 2.5.
With a dynamic repetition rate we no longer need to discard data.
Now, each Clifford sequence begins after the previous one with the same delay; all Clifford sequences therefore have the same state preparation error, see Fig.~\ref{fig:restless_rb}(b).

We compare standard and restless RB using Qiskit Experiments~\cite{QiskitExperiments} and the data processor described in Sec.~\ref{sec:data_processing}.
We measure $11$ and $14$ sequences of Clifford gates with $N_c$ ranging from 1 to 5101 and from 1 to 196 for single- and two-qubit RB, respectively.
Each length $N_c$ is measured for ten different random realizations with $N=1024$ shots.
In restless RB there is a fixed $1\,\mu{\rm s}$ delay after each measurement.
Single-qubit RB is done on qubit 13 of \emph{ibmq\_sydney}. 
We measure an EPC of $0.035 \pm 0.001 \%$ and $0.035 \pm 0.001 \%$ for three independent standard and restless RB experiments, respectively, see Fig.~\ref{fig:restless_rb}(c) and Appendix~\ref{sec:rb_appendix}.
Two-qubit RB is done on qubits~8 and 11 of \emph{ibmq\_sydney} for which we measure an $\mathrm{EPC}$ of $1.703 \pm 0.031 \%$ and $1.780 \pm 0.037 \%$ for three independent standard and restless RB experiments, respectively, see Fig.~\ref{fig:restless_rb}(d).
We observe a lower $A$ coefficient of the restless RB curves with respect to the standard one, see Fig.~\ref{fig:restless_rb}(c)-(d) and Appendix~\ref{sec:rb_appendix}. 
We attribute this scaling to $T_1$-induced state preparation errors in the restless measurements which reduce the probability that the qubit measured in $\ket{1}$ will also be measured in $\ket{1}$ after the subsequent Clifford sequence.
Crucially, this state preparation error does not significantly affect the measured depolarizing parameter $\alpha$.
To further illustrate this we split the single-qubit restless RB shots into two sets depending on the initial state of the qubit and analyze each set independently.
The data with the qubit in $\ket{0}$ has an $A$ value of $0.485(5)$, i.e. almost identical to the $0.487(3)$ produced by the standard measurement, while the data with the qubit in $\ket{1}$, shown in Fig.~\ref{fig:restless_rb}(c), has an $A$ of $0.426(3)$.
Finally, to demonstrate the importance of the restless data processor we process the restless data with the standard data processing chain.
This results in the useless green curves in Fig.~\ref{fig:restless_rb}(c) and~(d).

We compute the restless speed-up with Eq.~(\ref{eq:speed_up}).
Our schedules have a $\langle\tau_\text{circ}\rangle$ of $55.87\,\mu{\rm s}$ and $51.81\,\mu{\rm s}$ for single- and two-qubit RB, respectively.
Restless measurements therefore result in a $5.1\times$ and $5.3\times$ speed-up over standard measurements on \emph{ibmq\_sydney} for single- and two-qubit RB, respectively, see Tab.~\ref{tab:rb_runtime}.
On systems such as \emph{ibmq\_montreal} that have a $50\,\mu{\rm s}$ delay these speed-ups become $1.9\times$ and $1.7\times$ which emphasises the impact of the delay~\cite{Wack2021}.
We show RB data  in Appendix~\ref{sec:rb_appendix} taken on \emph{ibmq\_montreal} for different RB circuits.
The $T_1$ and $T_2$ times and readout errors of all used qubits can be found in Tab.~\ref{tab:devices} in Appendix~\ref{sec:appendix_qubits}.

\begin{table}[htbp!]
    \caption{
    Runtime breakdown on the quantum processor of RB with 11 and 14 different lengths of ten random Clifford gate sequences at each length for the one- and two-qubit experiment, respectively, with $N=1024$ shots. Therefore, $KN=10\cdot11\cdot1024$ and $KN=10\cdot14\cdot1024$.
    }
    \centering
    \begin{tabular}{l r c r r r}\hline\hline
         Processor / Restless & $\tau_\text{meas}$ & $\tau_\text{reset}^\dagger$ & $\tau_\text{delay}$ & $\langle{\tau}_\text{circ}\rangle$ & $\tau^{(x)}$\\
         & ($\mu {\rm s}$) & ($\mu {\rm s}$) & ($\mu {\rm s}$) & ($\mu {\rm s}$) & (${\rm s}$) \\
         \multicolumn{6}{l}{Single-qubit} \\ \hline
        \emph{ibmq\_sydney} / \ding{55} & 5.4 & 4 & 250.0 & 55.87 & 35.51\\
        \emph{ibmq\_sydney} / \ding{51} & 5.4 & n.m. & 1.0 & 55.87 & 7.01\\
        \emph{ibmq\_montreal} / \ding{55} & 5.2 & 4 & 50.0 & 55.87 & 12.96 \\
        \emph{ibmq\_montreal} / \ding{51} & 5.2 & n.m. & 0.5 & 55.87 & 6.93 \\ \\
         \multicolumn{6}{l}{Two-qubit} \\ \hline
        \emph{ibmq\_sydney} / \ding{55} & 5.4 & 4 & 250.0 & 51.81 & 44.61 \\
        \emph{ibmq\_sydney} / \ding{51} & 5.4 & n.m. & 1.0 & 51.81 & 8.34 \\
        \emph{ibmq\_montreal} / \ding{55} & 5.2 & 4 & 50.0 & 65.64 & 17.90 \\
        \emph{ibmq\_montreal} / \ding{51} & 5.2 & n.m. & 0.5 & 65.64 & 10.23 \\
          \hline
         \multicolumn{6}{l}{\footnotesize $\dagger$ We assume $4\,\mu{\rm s}$ since the backends do not disclose the exact} \\
         \multicolumn{6}{l}{\footnotesize duration of the reset which is typically between $3$ and $5\,\mu\rm{s}$.} \\
    \hline\hline
    \end{tabular}
    \label{tab:rb_runtime}
\end{table}

\section{Calibration\label{sec:cal}}

Calibration experiments determine the parameters of the control pulses to ensure quality.
Typically, different experiments are run where each is dedicated to measure a single parameter.
Here, we show a restless calibration workflow built on the data processor of Sec.~\ref{sec:data_processing}.
We run the calibration experiments on different devices based on system availability.
We illustrate restless calibration by calibrating DRAG pulses whose pulse-evelope is $\Omega_x(t)+i\beta \partial_t\Omega_x(t)$.
Here, $\Omega_x$ is a Gaussian pulse envelope with amplitude $A$ measured as a fraction of the maximum output of the arbitrary waveform generator. $\beta$ is the DRAG parameter~\cite{Motzoi2009}.

\subsection{Parameter scans\label{sec:param_scan}}

\begin{figure}[htbp!]
    \centering
    \includegraphics[width=1\columnwidth]{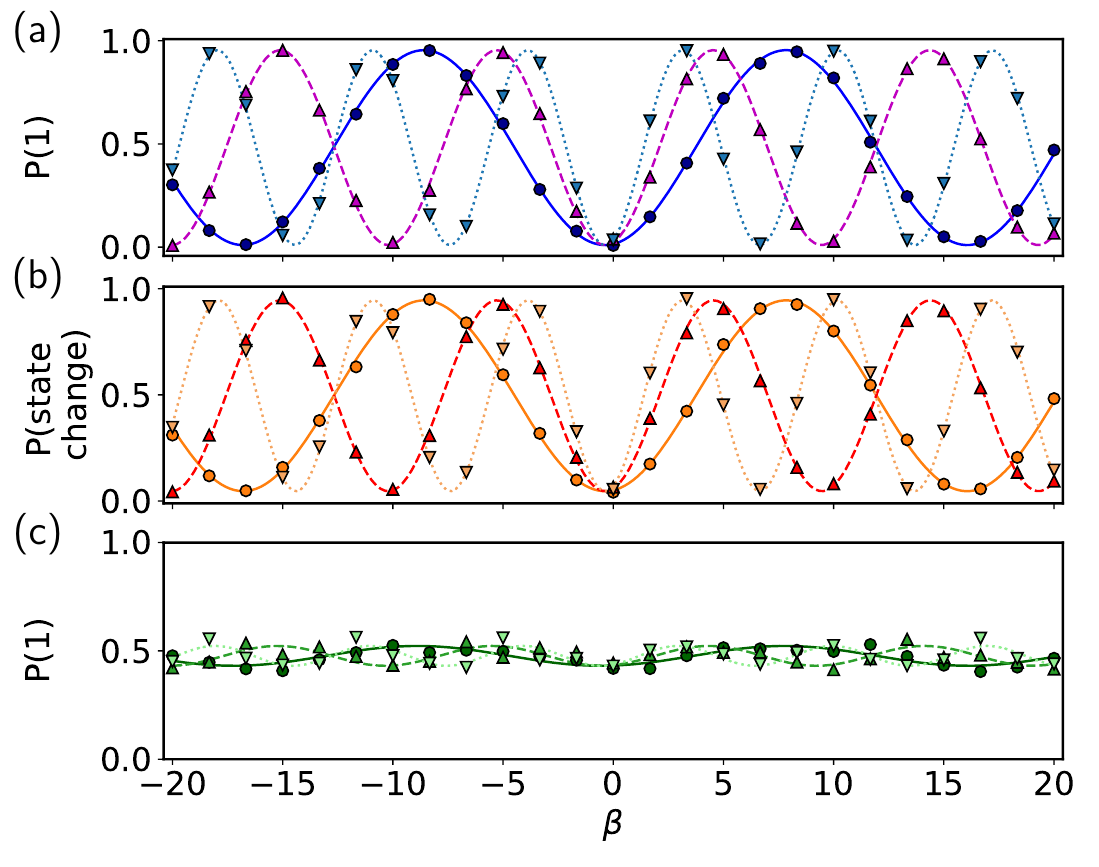}
    \caption{
    Rough DRAG calibration of \emph{ibmq\_quito} qubit~1.
    The solid, dashed, and dotted lines represent $n=3$, $5$, and $7$, respectively.
    (a) Standard measurements ($\chi^2 = 1.2$).
    (b) Restless measurements ($\chi^2 = 1.3$).
    (c) Restless measurements with the standard data processing. 
    }
    \label{fig:rough_drag}
\end{figure}

Parameter scans give a first estimate of a parameter value.
For example, the Rabi oscillation measured in Ref.~\cite{Werninghaus2021} yields a rough estimation of the amplitude of $\Omega_x$ as a function of the target rotation angle.
We now demonstrate a restless parameter scan to estimate $\beta$ using the gate sequence $[R_z(\pi)\cdot X(\beta)\cdot R_z(\pi)\cdot X(\beta)]^n$.
This sequence is repeated for different $\beta$ values and $n\in\{3,5,7\}$.
The virtual $R_z(\pi)$ rotations change the $X$ gate from a $\pi$-rotation to a $-\pi$-rotation~\cite{McKay2017}.
Ideally, this gate sequence rotates between the poles of the Bloch sphere in the $YZ$-plane and composes to the identity.
However, as phase errors accumulate, due to the higher-levels of the transmon, the state vector drifts out of the $YZ$-plane and oscillations in the qubit population appear~\cite{Gambetta2011, Chen2016}.
The measured population is fit to oscillating functions to estimate the $\beta$ that minimizes errors.
Standard and restless measurements on \emph{ibmq\_quito} both produce the same oscillating pattern with high-quality fits as indicated by the low $\chi^2$ values, see Fig.~\ref{fig:rough_drag}(a) and (b), resulting in a $\beta$ of $-0.376 \pm 0.006$ and $-0.355 \pm 0.007$, respectively, averaged over three independent measurements.
As expected, applying the standard data processor to restless measurements yields a useless signal, see Fig.~\ref{fig:rough_drag}(c).
Crucially, the restless speed-up obtained following Sec.~\ref{sec:speed_up} is $38.4\times$.

\subsection{Single-qubit error amplifying gate sequences\label{sec:error_amp}}

We now consider error amplifying gate sequences which repeat a gate pattern $n$ times to amplify an error $\mathrm{d}\theta$ to $n \cdot \mathrm{d}\theta$.
Typically, the measured qubit population $P$ (or the state change probability in the restless case) is fit to
\begin{align}\label{eq:fine_amp_fit}
\frac{a}{2} \cdot \cos\left[n \cdot (\theta + \mathrm{d}\theta) - \phi\right] + b.
\end{align}
Here, the intended angle per gate $\theta$ and the phase $\phi$ are fixed by the values of $P$ for ideal gates.
However, $a$, $b$, and ${\rm d}\theta$ are fit parameters.

\begin{figure}[htbp!]
    \centering
    \includegraphics[width=1\columnwidth]{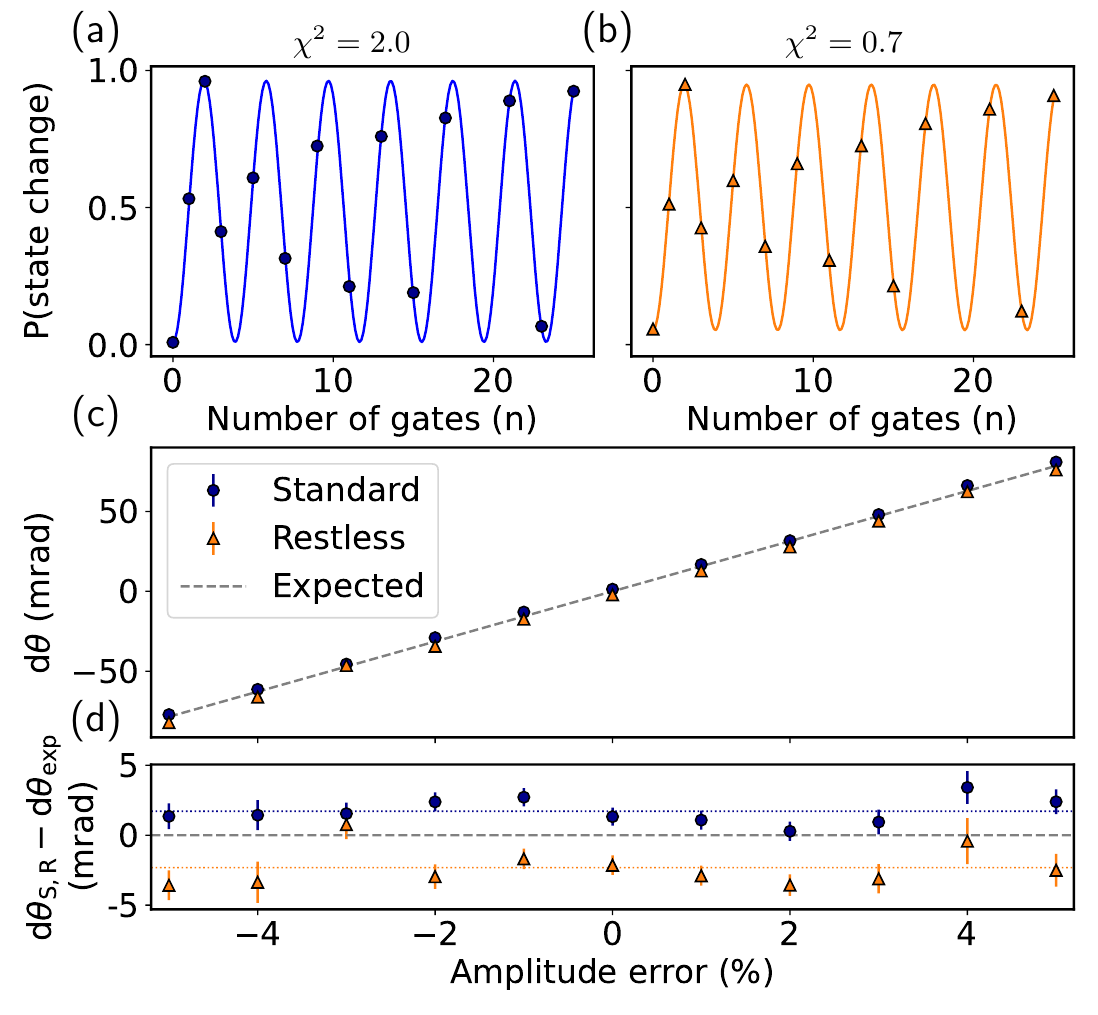}
    \caption{Fine amplitude calibration of a $\sqrt{X}$ gate on \emph{ibmq\_jakarta} qubit 1.
    Qubit population as a function of the number of $\sqrt{X}$ gates for standard (a) and restless measurements (b) with a $3\%$ amplitude error.
    (c) The measured deviation angle $\mathrm{d}\theta$ as a function of the added amplitude error.
    The dashed grey line indicates the expected deviation ${\rm d}\theta_\text{exp}$.
    (d) Deviation from the expected values ${\rm d}\theta_\text{exp}$.
    The subscripts S and R indicate the standard and restless data, respectively.}
    \label{fig:fine_sx_calibration}
\end{figure}

The pulse amplitude-error amplifying sequence of the $\sqrt{X}$ gate applies $\sqrt{X}$ an odd number of times. 
The ideal states lie on the equator of the Bloch sphere which maximizes the measurement's sensitivity to ${\rm d}\theta$.
The ideal gate sequence therefore imposes $\theta=\pi/2$ and $\phi= \pi$ in Eq.~(\ref{eq:fine_amp_fit}).
Two calibration points, one without any gates and one with two $\sqrt{X}$ gates, allow us to accurately fit $a$ in Eq.~(\ref{eq:fine_amp_fit}).
To demonstrate restless amplitude calibration we add an error ${\rm d}A$ to the amplitude $A$ of the $\sqrt{X}$ gate reported by qubit 1 of \emph{ibmq\_jakarta} ranging from $-5\%$ to $5\%$.
We interleave restless and standard jobs to mitigate biases due to drifts.
We measure the resulting ${\rm d}\theta$ by fitting the data obtained with both restless and standard measurements to Eq.~(\ref{eq:fine_amp_fit}). 
Both methods produce good fits as indicated by the low $\chi^2$ values, see Fig.~\ref{fig:fine_sx_calibration}(a) and (b).
The fitted rotation errors $\mathrm{d}\theta$ reproduce the expected deviation ${\rm d}\theta_\text{exp}=-\pi/2 \cdot {\rm d}A / ({\rm d}A + A)$ 
and on average differ by $1.7 \pm 0.9$\,mrad and $-2.3 \pm 1.3$\,mrad in the
standard and restless case, respectively, see Fig.~\ref{fig:fine_sx_calibration}(c) and (d).
As reference, note that a $1\,{\rm mrad}$ rotation error corresponds to a gate error of $2.5\cdot 10^{-7}$ on an otherwise perfect $\sqrt{X}$ gate.
We compute the restless speed-up for one experiment with Eq.~(\ref{eq:speed_up}).
With $\langle\tau_\text{circ}\rangle = 0.39\,\mu {\rm s}$, a default repetition delay of $250\,\mu {\rm s}$, and a restless repetition delay of $1\,\mu {\rm s}$ we obtain $\tau^{(s)} = 3.724\,{\rm s}$ and $\tau^{(r)} = 0.0973\,{\rm s}$ and therefore a $38.3 \times$ speed-up.

\begin{figure}[htbp!]
    \centering
    \includegraphics[width=1\columnwidth]{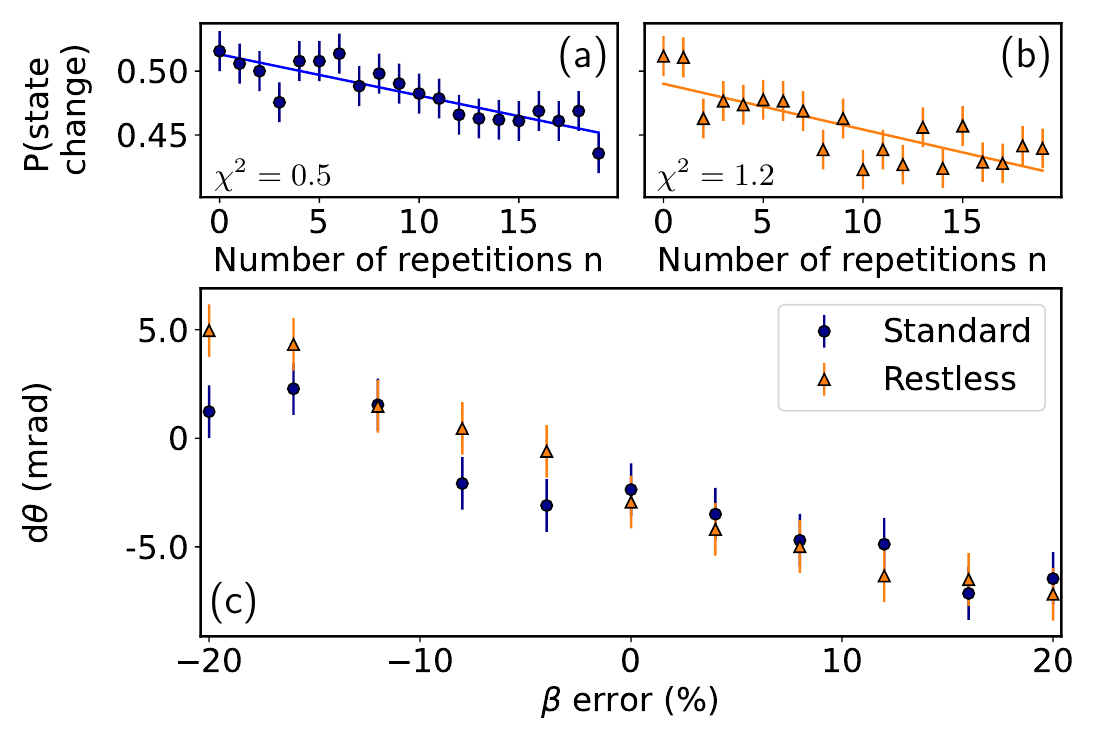}
    \caption{
    Characterization of $\beta$-induced errors on qubit 3 of \emph{ibmq\_montreal}.
    Qubit population as a function of the number of repetitions $n$ for standard (a) and restless measurements (b) with a $20\%$ DRAG parameter error.
    (c) The measured deviation angle $\mathrm{d}\theta$ as a function of the added $\beta$ error.}
    \label{fig:fine_sx_dragcal_montreal}
\end{figure}

We now consider the DRAG-error amplifying sequence $\sqrt{X}R_z(-\pi/2)\cdot[R_z(\pi)\cdot \sqrt{X}\cdot R_z(\pi)\cdot \sqrt{X}]^n\cdot \sqrt{X}$.
The first $\sqrt{X}$ rotation moves the state to the equator of the Bloch sphere.
The sequence $R_z(\pi)\cdot \sqrt{X}\cdot R_z(\pi)\cdot \sqrt{X}$ causes the state to oscillate between the equator and the $\ket{1}$ state.
The final $\sqrt{X}R_z(-\pi/2)$ gates map phase errors that took the state vector out of the $YZ$-plane during the oscillation onto the $Z$-axis for measurement.
We add an error ${\rm d}\beta$ ranging from $-20\%$ to $20\%$ to the calibrated $\beta$ value of the $\sqrt{X}$ gate reported by \emph{ibmq\_montreal} qubit~3.
Next, we measure DRAG-error amplifying sequences in a restless and standard setting.
The resulting data are fit to Eq.~(\ref{eq:fine_amp_fit}) with $\theta=0$, $\phi=\pi/2$ and $a$ fixed to 1 to measure ${\rm d}\theta$.
Once again, we observe a good agreement between standard and restless measurements, see Fig.~\ref{fig:fine_sx_dragcal_montreal}.
For this experiment the restless speed-up was $9.3\times$ since the default repetition delay of \emph{ibmq\_montreal} is $50\,\mu{\rm s}$.

\subsection{Two-qubit error amplifying gate sequences\label{sec:error_amp_2q}}

\begin{figure*}[htbp!]
    \centering
    \includegraphics[width=\textwidth]{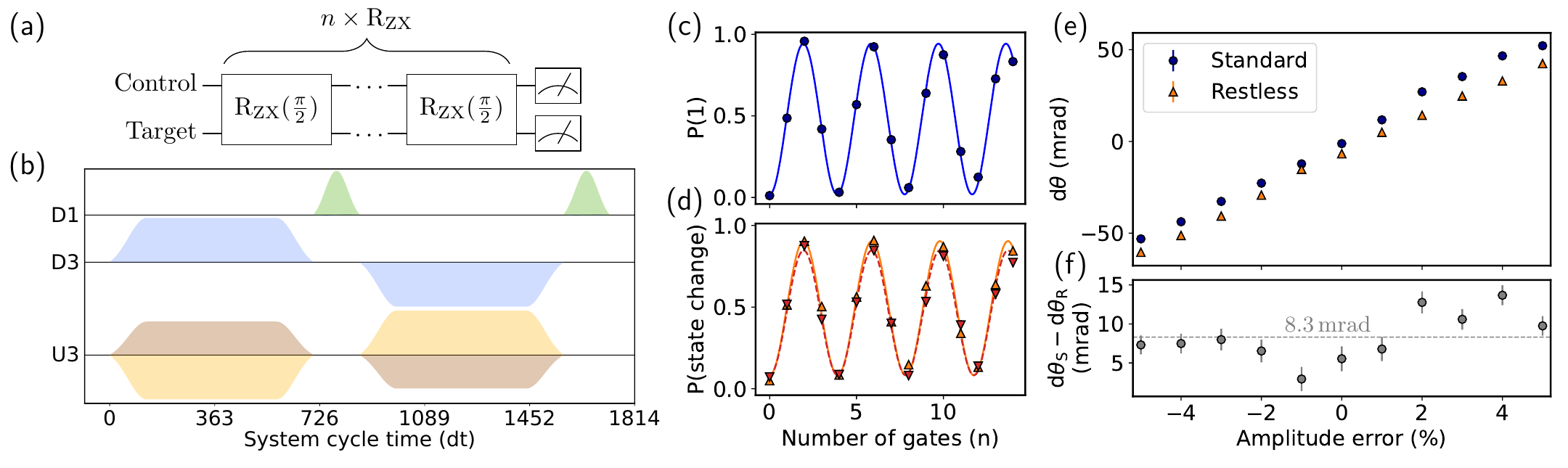}
    \caption{(a) Gate sequence to amplify amplitude errors of the $R_{ZX}$ gate. 
    (b) Echoed cross-resonance $R_{ZX}$ pulse schedule of \emph{ibmq\_jakarta} qubits 1 and 3.
    The $X$ pulses in the echo are shown in green on channel D1. 
    The rotary tones and cross-resonance drives are shown in blue and brown, respectively.
    The channel U3 is a control channel which applies pulses to the control qubit at the frequency of the target qubit.
    (c) Target qubit population of standard measurements with a 4\% amplitude error.
    (d) Probability of a state change in the target qubit obtained with restless measurements with a 4\% amplitude error.
    The solid and dashed lines show measurements where the control qubit was initialized in $\ket{0}$ and $\ket{1}$, respectively.
    (e) The measured deviation angle ${\rm d}\theta$ as a function of the added amplitude error. 
    The average size of the error bars on the standard and restless data are $1.4$ and $1.7~{\rm mrad}$, respectively.
    (f) Difference between the standard deviation angle ${\rm d}\theta_\text{S}$ and the restless deviation angle ${\rm d}\theta_\text{R}$ in~(e).}
    \label{fig:ecr_cal}
\end{figure*}

We now consider error amplifying gate sequences to determine amplitude errors for the two-qubit gate
\begin{align}
    R_{ZX}(\theta)=
    \begin{pmatrix}
    R_X(\theta) & 0 \\ 0 & R_X(-\theta)
    \end{pmatrix}
\end{align}
with restless measurements.
With an angle of $\theta=\pi/2$ this gate is a perfect entangler and can create a CNOT gate.
Fixed-frequency transmons implement $R_{ZX}$ with an echoed cross-resonance pulse sequence~\cite{Sheldon2016b} augmented with rotary tones~\cite{Sundaresan2020}.
Amplitude errors are amplified and measured by repetitively applying $R_{ZX}(\pi/2)$, see Fig.~\ref{fig:ecr_cal}(a).
We compare standard and restless measurements by adding an amplitude error ${\rm d}A$ to the CR pulses of the calibrated $R_{ZX}$ pulse schedule reported by the backend, see Fig.~\ref{fig:ecr_cal}(b).
When the qubits are reset only the target qubit needs to be measured which results in the same pattern as for the single-qubit $\sqrt{X}$ gate, compare Fig.~\ref{fig:fine_sx_calibration}(a) with Fig.~\ref{fig:ecr_cal}(c).
However, for restless measurements both qubits must be measured and some shots may project the control qubit, initially in the $\ket{0}$ state, into the $\ket{1}$ state.
This changes the sign of the subsequent rotations of the target qubit.
We account for this by computing the probability of a state change of the target qubit conditional on the control qubit.
This results in two data sets, shown in Fig.~\ref{fig:ecr_cal}(d), that exhibit the same pattern since an $R_X$ rotation with an angle $\pm n(\theta+{\rm d}\theta)$ results in the same qubit population when starting from the poles of the Bloch sphere.
We fit the data to two functions: both are given by Eq.~(\ref{eq:fine_amp_fit}) and share the same ${\rm d}\theta$ fit parameter.
However, each function has its own $a$ and $b$ parameters to accommodate differences in signal amplitude which we attribute to $T_1$ and the imperfect readout of the control qubit.
The restless measured ${\rm d}\theta$ closely follow the standard measurements, albeit with a $8.3 \pm 3.0\,{\rm mrad}$ bias which corresponds to a gate error of $1.7\cdot 10^{-5}$ on an otherwise perfect $R_{ZX}(\pi/2)$ gate.
For this experiment we obtain a restless speed-up of $28.8\times$.

\subsection{Calibration and characterization}

\begin{figure}[htbp!]
    \centering
    \includegraphics[width=1\columnwidth, clip, trim= 0 5 0 0]{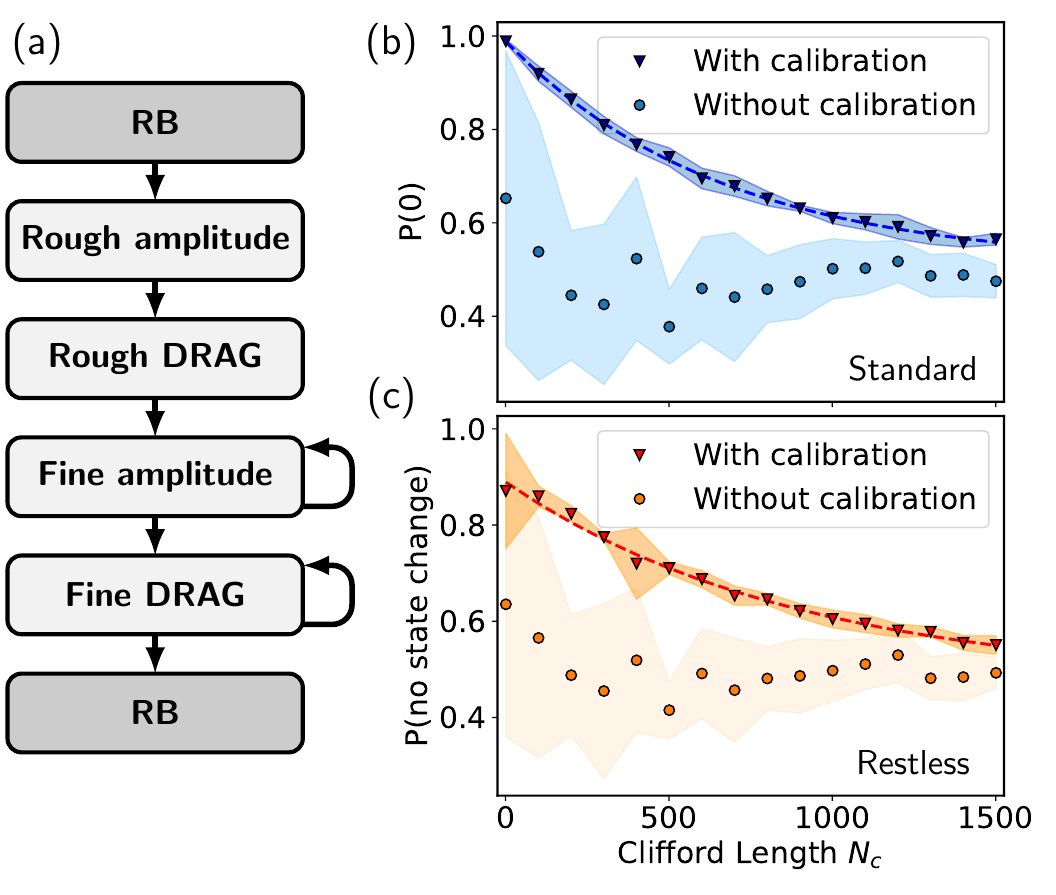}
    \caption{Calibration of $A$ and $\beta$ of a $\sqrt{X}$ gate. 
    (a) Schematic overview of the calibration experiments. 
    Parameter scans yield rough values for $A$ and $\beta$.
    Looped error-amplifying sequences are used as fine calibration experiments.
    Standard~(b) and restless (c) RB done before (light circles) and after (dark triangles) the calibration.
    }
    \label{fig:restless_sx_cal}
\end{figure}

We now tie sections~\ref{sec:rb}, \ref{sec:param_scan}, and \ref{sec:error_amp} together by calibrating and characterizing the $\sqrt{X}$ gate of \emph{ibmq\_bogota} qubit 2 with standard and restless measurements.
We calibrate the amplitude $A$ and DRAG-parameter $\beta$ and fix the duration and standard deviation of the Gaussian pulse to 160 and 40 samples, respectively. 
Each sample has a $0.222~{\rm ns}$ duration.
First, we measure a Rabi oscillation by scanning $A$ which yields a rough estimate of the amplitude~\cite{Werninghaus2021}.
Next, we scan $\beta$ using the experiment described in Sec.~\ref{sec:param_scan} which yields a rough estimate of the DRAG parameter.
We refine these rough estimates by looping the error-amplifying sequences described in Sec.~\ref{sec:error_amp}.
Here, each experiment is repeated until the measured ${\rm d}\theta$ is below a given tolerance or a maximum number of iterations is exceeded.
Prior to and after the calibration we measure the quality of the $\sqrt{X}$ gate with RB.
This workflow is shown in Fig.~\ref{fig:restless_sx_cal}(a).

As initial guess we chose $A = 0.25$ and $\beta = 0$.
With these values, both standard and restless RB show a low gate quality,
see the light blue and orange circles in Fig.~\ref{fig:restless_sx_cal}(b) and (c), respectively.
The amplitude and DRAG parameter values obtained from the rough calibrations are starting points for the iterative fine calibration experiments.
Both the standard and restless fine amplitude calibration experiments reach ${\rm d}\theta\leq 1\,{\rm mrad}$ tolerance  after five iterations while the fine DRAG calibration experiments required eight iterations.
The standard and restless calibration experiments yielded an amplitude of $A_S = (9.486 \pm 0.004) \cdot 10^{-2}$ and $A_R = (9.512 \pm 0.004) \cdot 10^{-2}$ and a DRAG parameter of $\beta_S = -0.39 \pm 0.03$ and $\beta_R = -0.29 \pm 0.03$, respectively. 
The backend reports $A = (9.501 + 0.080i) \cdot 10^{-2}$ and $\beta = -0.42$ as calibrated values.
Standard and restless RB of the pulses we calibrated produced an EPC of $0.075 \pm 0.006 \%$ and $0.066 \pm 0.013 \%$, respectively, see the dark blue and red triangles in Fig.~\ref{fig:restless_sx_cal}(b) and (c). 
The backend reported an error per $\sqrt{X}$ gate of $0.037 \%$.
The total execution time on the quantum processor, which does not include data-transfer, queuing and run-time compilation times, of the two RB and 15 calibration experiments was $\SI{187.0}{\second}$ and $\SI{11.6}{\second}$ for the standard and restless experiments, respectively. Here, the 15 calibration experiments accounted for $\SI{94.6}{\second}$ and $\SI{2.3}{\second}$ for standard and restless measurements, respectively.

\section{Process tomography\label{sec:qpt}}

Since RB is robust to SPAM errors restless and standard measurements produce identical results.
We now discuss restless process tomography measurements to characterize a quantum channel $\mathcal{E}$. 
Applying $\mathcal{E}$ on a density matrix $\rho^\text{in}_j$ results in the output density matrix
\begin{align}\label{eqn:qpt1}
    \rho^\text{out}_j=\mathcal{E}(\rho^\text{in}_j)=\sum_{m,n=0}^{d^2-1}\chi_{mn}E_m\rho^\text{in}_jE_n^\dagger.
\end{align}
Here, the $E_m$ form a basis of the $d\times d$ complex matrices where $d$ is the dimension of the Hilbert space.
Standard quantum process tomography reconstructs the matrix $\chi_{mn}$ describing $\mathcal{E}$ by preparing different input states $\rho_j^\text{in}$ and measuring them in a complete basis~\cite{OBrien2004, Mohseni2008, Nielsen2010, Bialczak2010}.
By writing each $\rho_j^\text{out}$ and $\rho_j^\text{in}$ in a common basis $\rho_k$ of density matrices Eq.~(\ref{eqn:qpt1}) becomes
\begin{align}\label{eqn:qpt2}
    c_{jk}=\sum_{mn}\chi_{mn}B_{mnjk},
\end{align}
where $\rho_j^\text{out}=\sum_k c_{jk}\rho_k$ is determined with state tomography. 
$B_{mnjk}$ depends on $\rho_j^\text{in}$.
Inverting Eq.~(\ref{eqn:qpt2}) yields the process matrix $\chi$ which can be made physical with different methods~\cite{Smolin2012, OBrien2004, Pechal2020}.
A pre-rotation $U^\mathrm{pre}_j$ applied to the initial state $\ket{0}$ creates $\rho_j^\text{in}$.
A complete input basis is formed for each qubit by choosing the gate set $\{I, X, H, SH\}$ as pre-rotations to prepare the eigenstates $\{Z_p, Z_m, X_p, Y_p\}$, respectively.
Here, for example, $Z_p$ and $Z_m$ denote the eigenstates of the $Z$ operator with positive and negative eigenvalues, respectively.
Choosing post-rotations $U^\mathrm{post}_i$ from $\{I,H,H S^\dagger\}$ allows us to measure along the $Z$, $X$, and $Y$-axis, respectively.

\begin{figure}[htbp!]
    \centering
    \includegraphics[width=0.9\columnwidth]{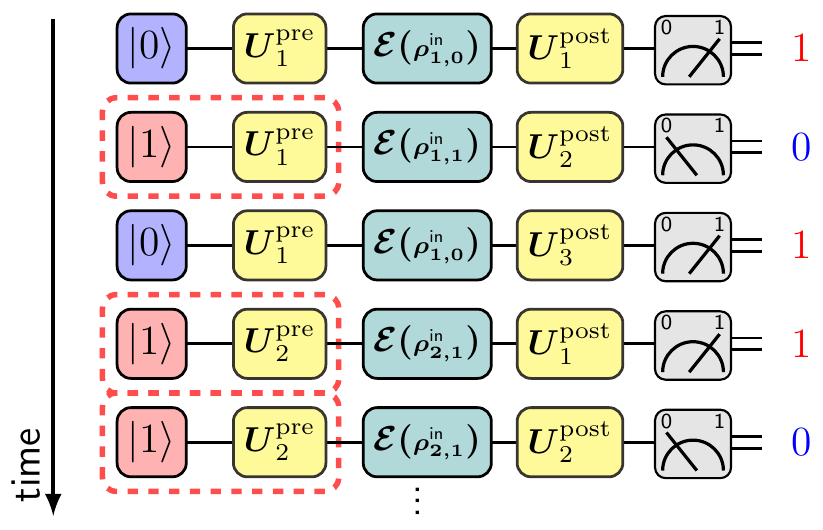}
    \caption{
    Illustration of the first five circuits of single-qubit restless QPT. 
    Each circuit consists of a pre-rotation $U^\mathrm{pre}_j$, the quantum process $\mathcal{E}$ to characterize, and a post-rotation $U^\mathrm{post}_i$.
    For the first circuit the qubit is initialized in the ground state.
    If the qubit is measured in the excited state the subsequent circuit starts with the qubit in state $\ket{1}$. 
    In a post-processing step these circuits are remapped to circuits with the pre-rotation $U^\mathrm{pre}_jX$, indicated by the dashed red boxes.
    }
    \label{fig:restless_qpt_concepts}
\end{figure}

\begin{figure}[htbp!]
    \centering
    \includegraphics[width=\columnwidth]{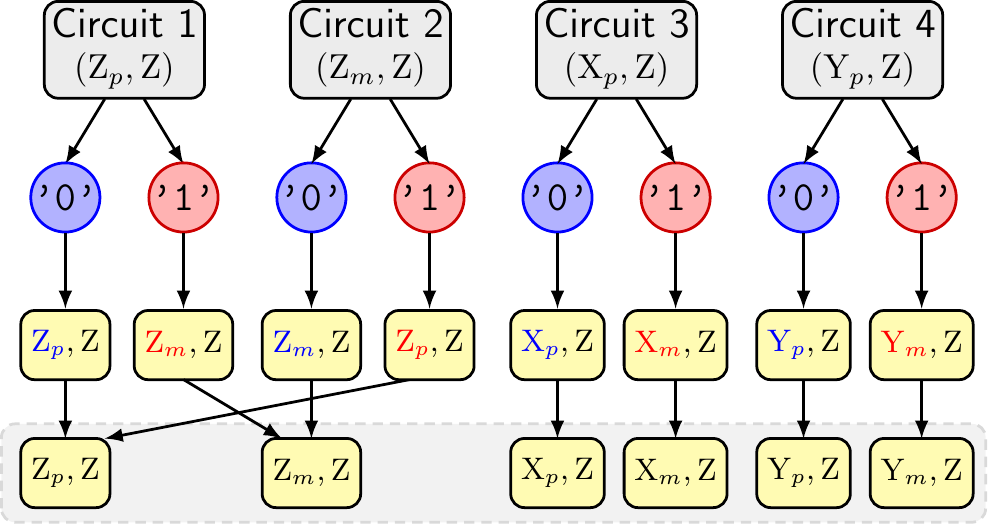}
    \caption{
    Ideal restless input state reassignment for $U_i^\text{post}=Z$.
    Since we time-order the measured outcomes we know the circuit that generated each shot (gray boxes) and the outcome of the previous measurement (circles).
    This allows us to reassign the measured outcome to the correct preparation basis (yellow boxes).
    Lastly, all outcomes are consolidated and assigned to the $\{Z_p, Z_m, X_p, X_m, Y_p, Y_m\}$ states.
    }
    \label{fig:restless_qpt_concepts_2}
\end{figure}

The input state of each qubit in an ideal restless measurement is either $\ket{0}$ or $\ket{1}$.
If the outcome of the previous measurement is $\ket{1}$, it is as if the pre-rotation is $U^\mathrm{pre}_jX$, see Fig.~\ref{fig:restless_qpt_concepts}. 
Therefore, in restless QPT, when the previous outcome was $\ket{1}$ the set of pre-rotations $\{I,X,H,SH\}$ is remapped to $\{X,I,HX, SHX\}$ which prepare the eigenstates $\{Z_m, Z_p, X_m, Y_m\}$, respectively.
A post-processing step is thus required to reassign the labels of the measured single-shots to the set of eigenstates $\{Z_p, Z_m, X_p, X_m, Y_p, Y_m\}$, see e.g. Fig.~\ref{fig:restless_qpt_concepts_2}.
We apply readout error mitigation to the resulting count dictionaries~\cite{Bravyi2020, Barron2020a}.
For one and two qubits this requires measuring two and four circuits to prepare each basis state.
Note, however, that this readout error mitigation does not overcome any restless-related state preparation errors.

\begin{figure}[htbp!]
    \centering
    \includegraphics[width=1\columnwidth]{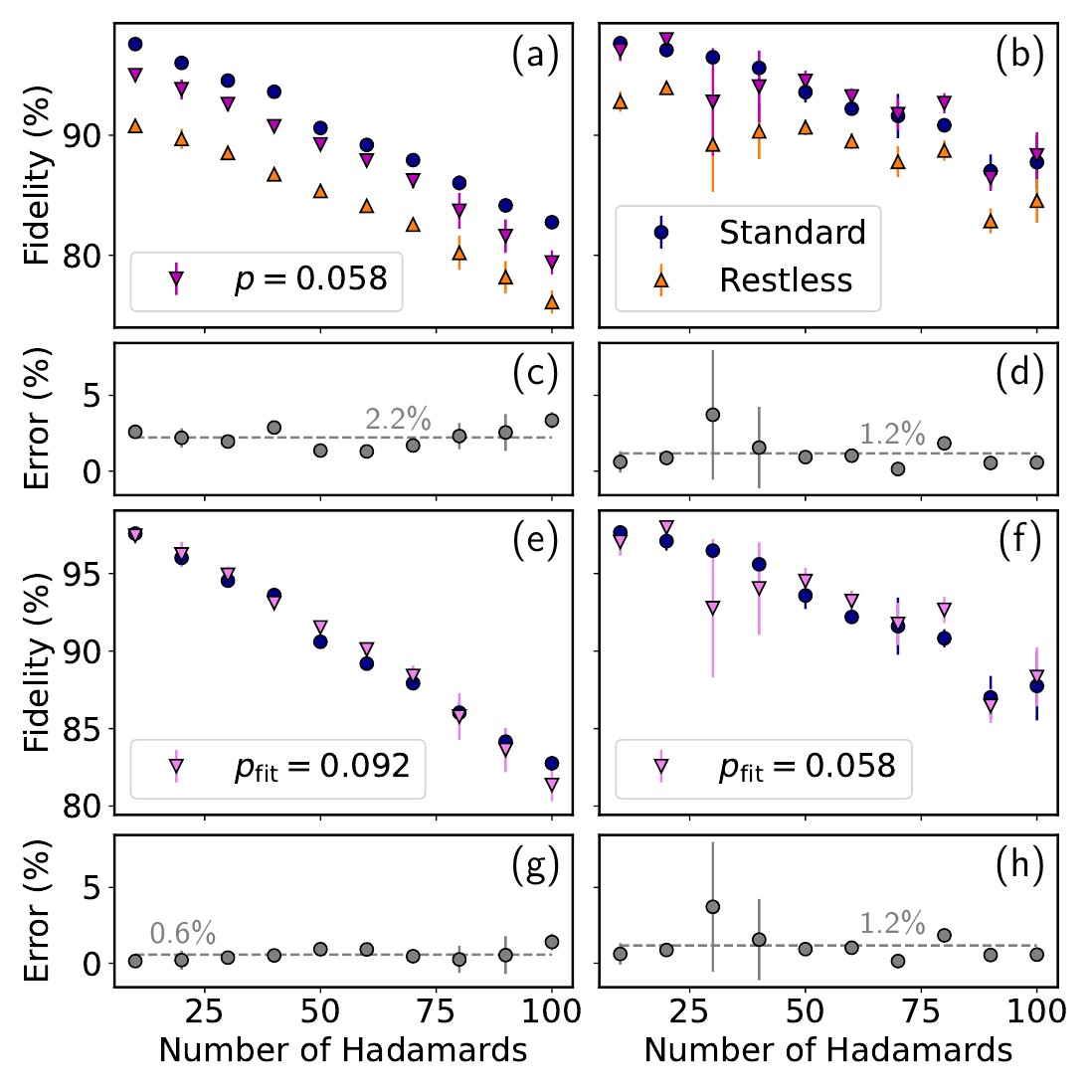}
    \caption{Standard and restless QPT of 10 to 100 Hadamard gates. The experiments were performed on \emph{ibmq\_sydney}, qubits 3 (a, c, d) and 5 (b, d, f). 
    Each fidelity is measured three times with 4096 shots; the markers show the average and the error bars the standard deviation.
    (a-b) Standard (blue dots), restless fidelities (orange up triangles) and restless fidelities computed using the preparation-error mitigated input states with $p$ based on $T_1$ (purple down triangles).
    (c-d) Difference between the purple down triangles and the blue circles.
    (e-f) Restless preparation-error mitigated fidelities based on a fit of the restless data to the standard QPT data (violet down triangles).
    (g-h) Difference between the data points in (e) and (f), respectively.}
    \label{fig:restless_qpt_hadamard_sydney}
\end{figure}

We benchmark the reconstruction process by performing standard and restless QPT of $n$ consecutive Hadamard gates with $n$ varying from 10 to 100 on \emph{ibmq\_sydney} qubits 3 and 5.
Standard QPT measures fidelities of $97.6 \pm 0.3 \%$ and $82.8 \pm 0.5 \%$ (qubit 3) and $97.7 \pm 0.2 \%$ and $87.8 \pm 2.2 \%$ (qubit 5) for 10 to 100 Hadamards, respectively.
We see in Fig.~\ref{fig:restless_qpt_hadamard_sydney}(a) and (b) that the fidelity $\mathcal{F}_R$ of the restless QPT is on average $6.0 \pm 0.6 \%$ and $3.9 \pm 1.4 \%$ lower than standard measured fidelity $\mathcal{F}_S$, respectively, compare the orange triangles with the blue circles.
This difference is independent of the process fidelity which opens up the possibility of using restless QPT as a fast cost function for optimal control.
This discrepancy is because measurement and $T_1$ may induce state preparation errors when the outcome $\ket{1}$ is measured.

\begin{figure}[htbp!]
    \centering
    \includegraphics[width=1\columnwidth]{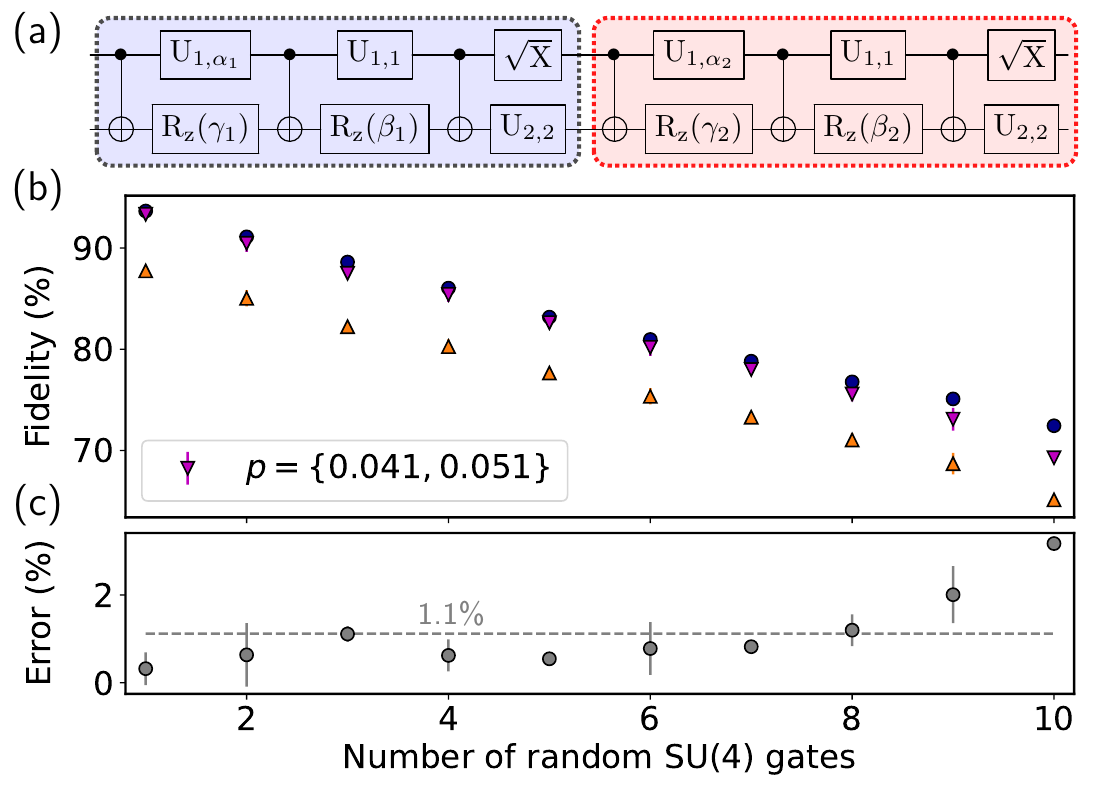}
    \caption{Standard and restless QPT of one to ten random SU(4) gates. The experiments were performed on \emph{ibmq\_sydney}, qubits 3 and 5. (a) Two SU(4) gates with angles ($\alpha_i$, $\beta_i$, $\gamma_i$) chosen at random in the Weyl chamber.
    (b) Standard QPT (blue dots), post-processed restless QPT (orange triangles) and restless preparation basis QPT results (purple triangles) for qubit 3 and 5.  
    Each fidelity is measured three times with 4096 shots.
    The $T_1$-induced decay probability is $p \approx 0.041$ and $p \approx 0.051$ for qubits 3 and 5, respectively.
    (c) Absolute deviation of the restless preparation basis QPT results from the standard QPT results.}
    \label{fig:restless_qpt_cnots_sydney}
\end{figure}

\subsection{Error mitigation}

So far we assumed that the state after each readout was either $\ket{0}$ or $\ket{1}$ since the measurement pushes the qubit towards eigenstates of the $Z$ operator~\cite{Gambetta2008} while increasing dephasing~\cite{Schuster2005, Gambetta2006}.
However, the finite qubit $T_1$ causes $\ket{1}\to\ket{0}$ jumps during the delay and the readout with probability $p$ which we approximate by $p = 1 - \exp{(-[\tau_{\rm meas} + \tau_{\rm delay}]/T_1)}$.
We therefore model the initial state following a $\ket{1}$ outcome by an amplitude damping channel $\mathcal{E}_\text{ad}$ with strength $p$ applied to $\ket{1}\!\!\bra{1}$, i.e. $\mathcal{E}_\text{ad}(\ket{1}\!\!\bra{1})=p\ket{0}\!\!\bra{0}+(1-p)\ket{1}\!\!\bra{1}$.
To mitigate the errors in restless QPT shown in Fig.~\ref{fig:restless_qpt_hadamard_sydney}(a) and (b) we modify the preparation basis which now includes the eight states
\begin{align}
\label{eq:restless_prep_basis_0}
    \rho_{j,0}^{\text{in}}=&U_j^\text{pre}\ket{0}\!\!\bra{0}(U_j^\text{pre})^\dagger\quad\text{and} \\ \label{eq:restless_prep_basis_1}
    \rho_{j,1}^{\text{in}}=&U_j^\text{pre}\left[p\ket{0}\!\!\bra{0}+(1-p)\ket{1}\!\!\bra{1}\right](U_j^\text{pre})^\dagger,
\end{align}
where $j\in\{1,2,3,4\}$, see details in Appendix~\ref{sec:appendix_qpt}.
The decay probability is $p = 0.058$ for both qubits 3 and 5 since  at the time the QPT data was taken they had a $T_1$ of $106.5\,\mu \mathrm{s}$ and $105.7\,\mu \mathrm{s}$, respectively, while $\tau_{\rm meas} = 5.4\,\mu {\rm s}$, and $\tau_{\rm delay}=1\,\mu{\rm s}$. 
With the modified preparation basis in Eq.~(\ref{eq:restless_prep_basis_0}) and (\ref{eq:restless_prep_basis_1}) the error between the restless and standard fidelities is reduced to $2.2 \pm 0.6 \%$ and $1.2 \pm 1.0 \%$ on average for qubits 3 and 5, respectively, see Fig.~\ref{fig:restless_qpt_hadamard_sydney}(c) and (d) and the purple triangles in Fig.~\ref{fig:restless_qpt_hadamard_sydney}(a) and (b).
This error is further reduced if we treat $p$ in Eq.~(\ref{eq:restless_prep_basis_1}) as a fit parameter.
We thus find the $p_\text{fit}$ that minimizes the sum of squared errors between the restless and standard fidelities, i.e. $p_\text{fit}=\min_p\sum [\mathcal{F}_R(p)-\mathcal{F}_S]^2$ where the sum is carried out over each QPT measurement.
This results in $p_{\rm fit} = 0.092$ and $p_{\rm fit} = 0.058$ for qubits 3 and 5, respectively, see Fig.~\ref{fig:restless_qpt_hadamard_sydney}(e) and (f).
This further reduces the error between restless and standard QPT measurements to $0.6 \pm 0.4 \%$ for qubit 3, on average, see Fig.~\ref{fig:restless_qpt_hadamard_sydney}(g) and (h). 
This shows that $p$ suffices to describe the discrepancy between restless and standard QPT data. 
If the effect of the measurement process is well understood we can compensate the $T_1$-induced errors using the restless preparation basis.
The fluctuations in the data for qubit 5 suggest that restless QPT is more unstable than standard QPT. 
This could be explained by the higher sensitivity to measurement or $T_1$-induced errors since $T_1$ is known to fluctuate~\cite{Klimov2018, Carroll2021}.

We further demonstrate two-qubit restless QPT of one to ten random SU(4) gates on qubits 3 and 5 of \emph{ibmq\_sydney}. Here, the $i$-th two-qubit gate is decomposed into three CNOT gates and single-qubit gates~\cite{Vidal2004}, see Fig.~\ref{fig:restless_qpt_cnots_sydney}(a).
The single-qubit gates depend on the three angles ($\alpha_i$, $\beta_i$, $\gamma_i$) that we chose at random within the Weyl chamber~\cite{Zhang2003, Khaneja2001, Byron2008}. 
We observe a high error of $6.0 \pm 0.5 \%$, on average, between $\mathcal{F}_\text{R}$ and $\mathcal{F}_\text{S}$ when $p=0$, see the orange and blue markers in Fig.~\ref{fig:restless_qpt_cnots_sydney}(b).
However, when we compute the restless fidelities with the decay probabilities $p = \{0.041, 0.051\}$, based on the $T_1$-times of $151.0~\mu{\rm s}$ for qubit 3 and $122.4~\mu{\rm s}$ for qubit 5 at the time the QPT data was taken, the average deviation between the restless fidelities and the standard fidelities is reduced to $1.1 \pm 0.8 \%$, see Fig.~\ref{fig:restless_qpt_cnots_sydney}(c).
Errors of up to 3.2\%, e.g. for ten random SU(4) gates, are still present which we attribute to the higher sensitivity of restless QPT to SPAM errors. 

We also run restless QPT on \emph{ibmq\_manila} and \emph{ibmq\_montreal}, the data, presented in Appendix~\ref{sec:appendix_qpt}, show a similar behaviour.
The $T_1$ and $T_2$ times and readout errors of all used qubits are listed in Tab.~\ref{tab:devices} in Appendix~\ref{sec:appendix_qubits}.

\subsection{Restless QPT speed-up}

We compute the restless speed-up using Eq.~(\ref{eq:speed_up}).
For single-qubit QPT with ten Hadamard gates and two-qubit QPT with one random SU(4) gate on \emph{ibmq\_sydney} we obtain an average circuit duration of $\langle\tau_\text{circ}\rangle$ of $0.41\,\mu {\rm s}$ and $1.24\,\mu {\rm s}$, respectively, see Tab.~\ref{tab:qpt_runtime}. 
A restless repetition delay of $1\,\mu {\rm s}$ leads to a $38.7\times$ and $34.1\times$ speed-up for single- and two-qubit QPT, respectively. 
On systems like \emph{ibmq\_montreal} with a default repetition delay of $50\,\mu {\rm s}$ and a restless repetition delay of $0.5\,\mu {\rm s}$ we obtain a $9.8 \times$ and $8.3 \times$ speed-up for single- and two-qubit QPT, respectively, see Appendix~\ref{sec:appendix_qpt}.

\begin{table}[htbp!]
    \caption{
    Runtime breakdown on the quantum processor of one- and two-qubit QPT of ten Hadamard gates and one random SU(4) gate with $N_\text{shots}=4096$.
    The single- and two qubit \emph{ibmq\_sydney} QPT data were measured on November 2, 2021 and December 21, 2021, respectively.
    }
    \centering
    \begin{tabular}{l c c r c r}\hline\hline
         Processor / Restless & $\tau_\text{meas}$ & $\tau_\text{reset}^\dagger$ & $\tau_\text{delay}$ & $\langle{\tau}_\text{circ}\rangle$ & $\tau^{(x)}$\\
         & ($\mu {\rm s}$) & ($\mu {\rm s}$) & ($\mu {\rm s}$) & ($\mu {\rm s}$) & (${\rm s}$) \\ 
         \multicolumn{6}{l}{Single-qubit QPT} \\ \hline
         \emph{ibmq\_sydney} / \ding{55} & 5.4 & 4 & 250.0 & 0.41 & 12.77 \\
         \emph{ibmq\_sydney} / \ding{51} & 5.4 & n.m. & 1.0 & 0.41 & 0.33 \\
         \emph{ibmq\_montreal} / \ding{55} & 5.2 & 4 & 50.0 & 0.41 & 2.93 \\
         \emph{ibmq\_montreal} / \ding{51} & 5.2 & n.m. & 0.5 & 0.41 & 0.30 \\ \\
         \multicolumn{6}{l}{Two-qubit QPT} \\ \hline
         \emph{ibmq\_sydney} / \ding{55} & 5.4 & 4 & 250.0 & 1.24 & 153.73 \\
         \emph{ibmq\_sydney} / \ding{51} & 5.4 & n.m. & 1.0 & 1.24 & 4.51 \\
         \emph{ibmq\_montreal} / \ding{55} & 5.2 & 4 & 50.0 & 1.64 & 35.89 \\
         \emph{ibmq\_montreal} / \ding{51} & 5.2 & n.m. & 0.5 & 1.64 & 4.33 \\
          \hline
         \multicolumn{6}{l}{\footnotesize $\dagger$ We assume $4\,\mu{\rm s}$ since the backends do not disclose the exact} \\
         \multicolumn{6}{l}{\footnotesize duration of the reset which is typically between $3$ and $5\,\mu\rm{s}$.} \\
    \hline\hline
    \end{tabular}
    \label{tab:qpt_runtime}
\end{table}

\section{Conclusion\label{sec:conclusion}}

We have investigated restless calibration and characterization on cloud-based quantum computers.
Turning active reset off and reducing the delay between the end of a measurement and the beginning of the next circuit creates a restless setting.
Crucially, the restless measurements are now done with a dynamic repetition rate as opposed to a fixed repetition rate~\cite{Werninghaus2021}.
This enables randomized benchmarking with excellent agreement to standard measurements without discarding any data as done in Ref.~\cite{Werninghaus2021} which also limited the scalability of restless measurements.

We have also demonstrated restless qubit calibration of the amplitude and DRAG parameter of a single-qubit pulse.
Here, restless measurements yield a speed-up ranging from $9\times$ to $38\times$.
When including the RB characterization the qubit calibration is $16\times$ faster with restless measurements.
These speed-ups depend on the quality of the active-reset to which we compare.
As active reset improves the default repetition delay on the system will shorten, therefore lowering the speed-up.
Restless measurements will nevertheless have the lowest run-time on the quantum processor since they require the lowest number of operations.
Restless measurements have the added benefit that they do not require active reset therefore eliminating reset-related issues from fast characterization and calibration tasks.
Importantly, the speed-ups that we measure do not take into account the classical run-time compilation and data transfer times which can be significant~\cite{Wack2021}.
This shows that increasing the speed at which quantum systems are calibrated depends both on the quantum and the classical hardware.
Here, we have focused on the quantum hardware.

Furthermore, we have demonstrated restless measurements of a two-qubit error amplifying gate sequence.
Two-qubit experiments have the added complexity that the effect of the gate depends on the state of both qubits.
The restless post processing must accommodate this by possibly separating the data in two series conditional on the state of the control qubit.

We have also shown that it is possible to perform QPT with restless measurements as long as the effect of the readout is well understood and the system is stable.
Speeding-up QPT becomes even more relevant as the system size increases~\cite{Galda2021}.
This task may be complicated by any effect that the readout may have on the qubit~\cite{Goy1983, Krantz2019}.
We observed that restless QPT measured fidelities scale with the number of gates as expected even when state preparation errors are not mitigated.
The $10\times$ to $39\times$ speed-up afforded by restless QPT opens up the possibility to use it as a fast cost function for closed-loop optimal control~\cite{Kelly2014, Egger2014} even if does not exactly match the fidelity with standard measurements.
By contrast to QPT, gate set tomography~\cite{Merkel2013, Blumekohout2013, Greenbaum2015, BlumeKohout2017} is less vulnerable to state preparation and measurement errors.
Future work may therefore explore whether gate set tomography can be performed with restless measurements and extend restless measurements to many-qubit protocols such as Cycle Benchmarking which measures the performance of parallel gate cycles and is robust to SPAM errors~\cite{Erhard2019}.

In conclusion, restless measurements offer a simple and fast alternative to reset-based measurements for calibration and characterization tasks~\cite{Ganzhorn2020}.
This can increase the availability of cloud-based systems by reducing calibration and characterization time or increase quality by running calibration more often.
Restless measurements may also make it easier to handle the large calibration overhead required by Richardson error mitigation which calibrates several scaled versions of the same pulse set~\cite{Temme2017, Kandala2019}.
Finally, many error correcting protocols, such as the surface code~\cite{Fowler2012, Chamberland2020}, require repeated measurements of syndrome qubits.
We envision that restless measurements may be used in error correcting codes to speed-up the error detection cycle -- a key driver of the speed of fault-tolerant quantum computers -- this may help obtain a quantum advantage at smaller problem sizes~\cite{Chakrabarti2021}.

\subsection{Acknowledgements}
The authors acknowledge A. Wack and O. Dial for helpful discussions and the Qiskit Experiments development team.
We acknowledge the use of IBM Quantum services
for this work.
The views expressed are those of the authors, and do not reflect the official policy or position of IBM or the IBM Quantum team.

IBM, the IBM logo, and ibm.com are trademarks of International Business Machines Corp., registered in many jurisdictions worldwide.
Other product and service names might be trademarks
of IBM or other companies.
The current list of IBM trademarks is available at https: //www.ibm.com/legal/copytrade.

\appendix

\section{Dynamic repetition delay\label{sec:restless_delay}}

\begin{figure}[htbp!]
    \centering
    \includegraphics[width=\columnwidth]{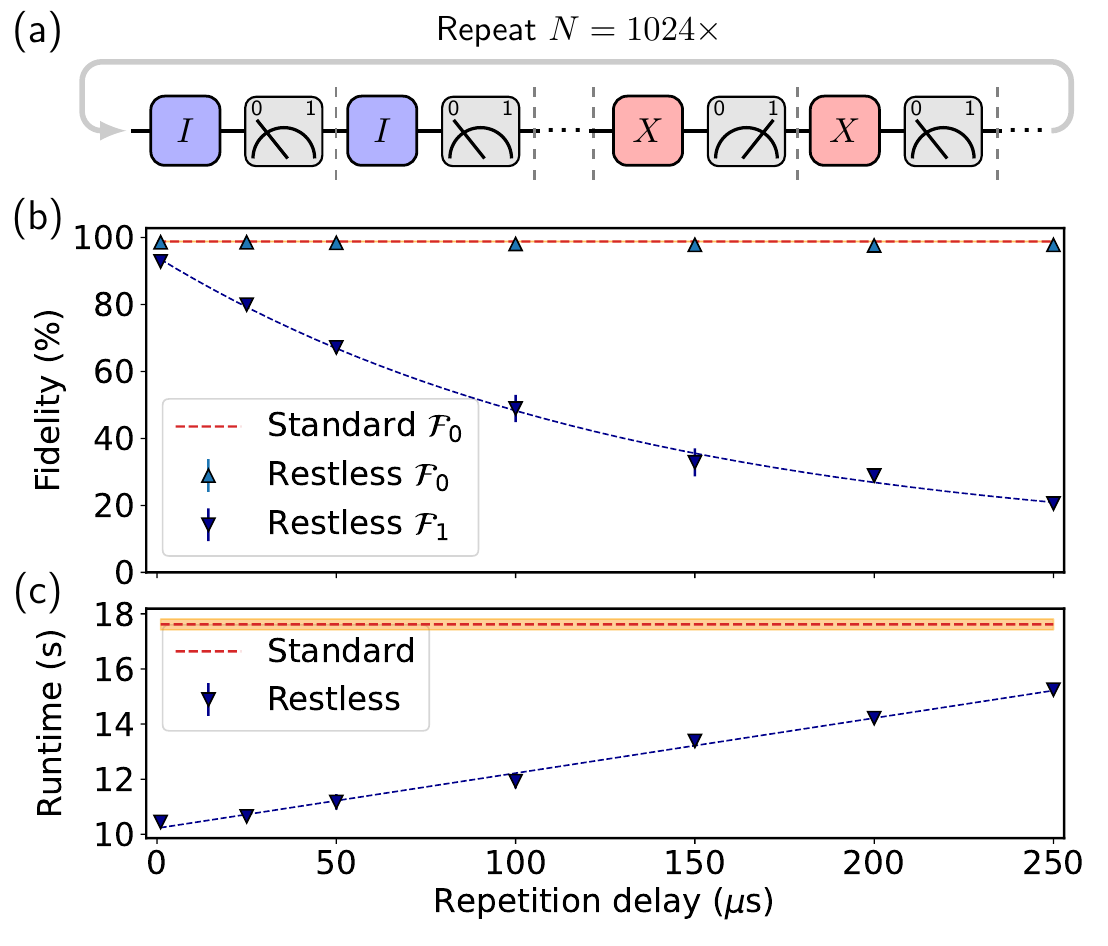}
    \caption{
    Restless state preparation and measurement fidelities and total runtimes on \emph{ibmq\_sydney} qubit 5 for different repetition delays. 
    (a) Twenty circuits with 10 identity and 10 $X$ gates are measured. 
    (b) Restless measurement fidelities $\mathcal{F}_0$ (blue up-triangles) and $\mathcal{F}_1$ (dark blue down-triangles). 
    Each point is the average of three individual measurements. 
    The red dashed line shows the average standard fidelity $\mathcal{F}_0$ computed from the results of $3 \times 7$ interleaved jobs with active reset.
    (b) Mean reported total runtimes of the circuits in (b) in the standard (red dashed line) and restless setting (dark blue down-triangles).
    The shaded orange area and the error bars correspond to the standard deviation of the standard and restless runtimes, respectively.}
    \label{fig:restless_readout}
\end{figure}

To illustrate the impact of the repetition delay on restless measurements we compute the state preparation and measurement fidelity as in Ref.~\cite{Werninghaus2021}.
We measure 20 circuits; the first ten are an ${I}$ gate followed by a measurement and the second ten are an ${X}$ gate followed by a measurement, see Fig.~\ref{fig:restless_readout}(a).
With these circuits we measure two distinct errors.
The first error occurs when circuit $k-1$ measures $\ket{0}$ but the outcome of circuit $k$ is $\ket{1}$ for an ${I}$ gate or $\ket{0}$ for an ${X}$ gate, i.e. $P_0(1|I) + P_0(0|X)$.
The second error occurs when circuit $k-1$ measures $\ket{1}$ but the outcome of circuit $k$ is $\ket{1}$ for an ${X}$ gate or $\ket{0}$ for an ${I}$ gate, i.e. $P_1(0|I) + P_1(1|X)$.
From these errors we compute the state preparation and measurement fidelities 
\begin{align}
    \mathcal{F}_0 &= 1 - \frac{1}{2}\left[P_0(1|I) + P_0(0|X)\right],\quad\text{and} \\
    \mathcal{F}_1 &= 1 - \frac{1}{2}\left[P_1(0|I) + P_1(1|X)\right].
\end{align}
We evaluate these fidelities for different repetition delays ranging from $1\,\mu{\rm s}$ to $250\,\mu{\rm s}$ and with 1024 shots.
At $1\,\mu{\rm s}$ we measure $\mathcal{F}_0=98.57\pm 0.03\%$ and $\mathcal{F}_1=92.86\pm 0.57\%$.
We observe that $\mathcal{F}_0$ is independent of the repetition delay while $\mathcal{F}_1$ decays exponentially, see Fig.~\ref{fig:restless_readout}(b). 
This decay is caused by $T_1$ as confirmed by the exponential fit to $\mathcal{F}_1$ resulting in $T_{1, {\rm fit}} = 131.5\,\mu {\rm s}$ in close agreement with the $134.1\,\mu {\rm s}~T_1$ reported by the backend. 
As a reference, we also run the circuits with an active reset and the default repetition delay of $250\,\mu {\rm s}$ interleaved and calculate $\mathcal{F}_0$. 
The resulting fidelities agree with $\mathcal{F}_0$ measured in the restless setting, see the orange dashed line in Fig.~\ref{fig:restless_readout}(b), which is the average over 21 independent measurements.

IBM Quantum systems also report job execution times as the sum of the quantum processor time and the classical compile and data transfer times.
The execution time when active reset is used with a repetition delay of $250\,\mu{\rm s}$ is constant at $17.6\pm 0.2~{\rm s}$ averaged over 21 independent jobs, see the dashed line in Fig.~\ref{fig:restless_readout}(c).
For restless measurements the execution time increases linearly as a function of the repetition delay.
The measured slope is $0.02~{\rm s}/\mu{\rm s}$ which corresponds to an extra $20\times 1024\times 10^{-6} \approx 0.02$ second per each extra $\mu{\rm s}$ of added delay as expected from Eq.~(\ref{eq:speed_up}) and the 20 circuits each measured 1024 times.

\section{Additional Restless Randomized Benchmarking data\label{sec:rb_appendix}}

\begin{figure}[htbp!]
    \centering
    \includegraphics[width=1\columnwidth]{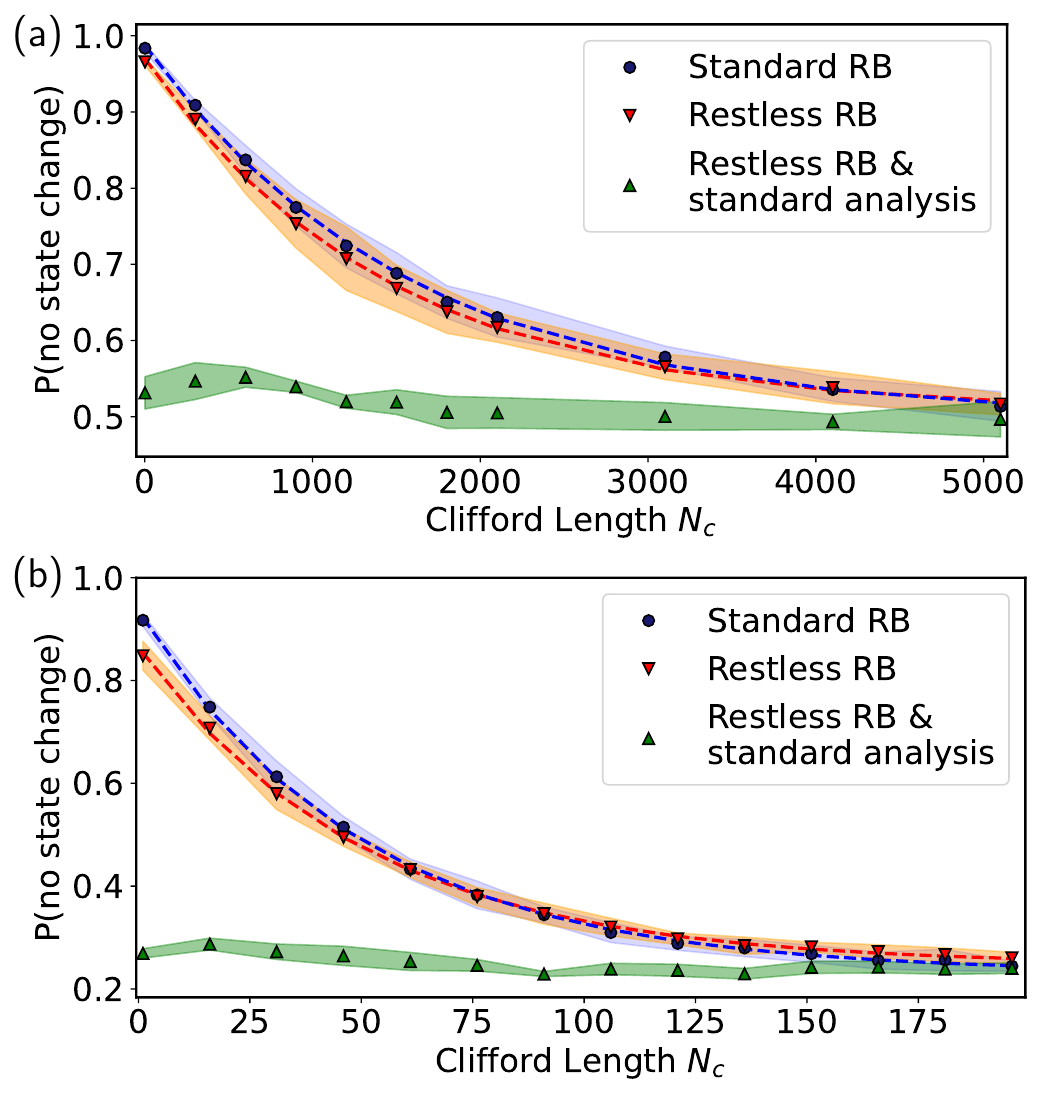}
    \caption{Standard and restless RB on \emph{ibmq\_montreal}.
    (a) Single-qubit RB for qubit 0. 
    (b) Two-qubit RB for qubits 1 and 2.
    Each marker indicates the mean value of ten Clifford sequences and the shaded area shows the standard deviation.
    The green triangles show restless data analyzed with the standard data processor.
    }
    \label{fig:restless_rb_montreal}
\end{figure}

Here, we show in Tab.~\ref{tab:RB} the fit parameter $\alpha$ and the EPC for each of the three RB measurements done on \emph{ibmq\_sydney} which were summarized in the main text.
We also present additional restless and standard RB data acquired on \emph{ibmq\_montreal}.
On \emph{ibmq\_montreal} standard measurements have a default $50\,\mu{\rm s}$ delay after a reset.
We perform restless RB with a $0.5\,\mu{\rm s}$ delay after each measurement.
Single-qubit RB on qubit 0 yields an $\mathrm{EPC} = 0.031 \pm 0.001 \%$ and $\mathrm{EPC} = 0.033 \pm 0.001 \%$ for three independent standard and restless RB experiments, respectively, see Fig.~\ref{fig:restless_rb_montreal} and Tab.~\ref{tab:RB}.

Two-qubit RB is done on qubits 1 and 2 of \emph{ibmq\_montreal} for which we measure $\mathrm{EPC} = 1.502 \pm 0.027 \%$ and $\mathrm{EPC} = 1.495 \pm 0.035 \%$ for three independent standard and restless RB experiments, respectively, see Fig.~\ref{fig:restless_rb_montreal}(b).
To demonstrate the importance of the restless data processor we process the restless data with the standard data processing chain in which the measured outcomes are simply aggregated in a counts dictionaries.
This results in the useless green curve in Fig.~\ref{fig:restless_rb_montreal}(a) and (b).

\begin{table}[htbp!]
    \centering
    \caption{
    Standard and restless RB fit data.
    $N_c^\text{max}$ indicates the maximum number of Clifford gates in an RB experiment.
    The errors indicate one standard deviation.
    Sydney and Montreal stand for the \emph{ibmq\_sydney} and \emph{ibmq\_montreal} systems, respectively.}
    \label{tab:RB}
    \begin{tabular}{l c c c c}
        \hline\hline
        Experiment & $\alpha$ & EPC & $A$ & $B$ \\
        & $(\%)$ & $(\%)$ & & \\ 
         \hline
         \emph{ibmq\_sydney} &  &  &  &    \\
         qubit 13 & 99.945(1) & 0.028(1) & 0.493(4) & 0.503(4) \\
         standard & 99.946(1) & 0.027(1) & 0.494(6) & 0.501(6) \\
         $N_c^\text{max}=5101$  & 99.901(2) & 0.050(1) & 0.487(3) & 0.505(3) \\
         \hline
        \emph{ibmq\_sydney} &  &  &  &    \\
        qubit 13 & 99.945(2) & 0.028(1) & 0.475(6) & 0.493(6) \\
        restless & 99.941(2) & 0.029(1) & 0.463(5) & 0.503(6)\\
        $N_c^\text{max}=5101$ & 99.902(2) & 0.049(1) & 0.458(4) & 0.496(3)  \\
         \hline
        \emph{ibmq\_sydney} &  &  &  &    \\
        qubits 8 \& 11 & 97.745(42) & 1.692(32) & 0.669(4) & 0.259(3) \\
        standard & 97.709(38) & 1.718(28) &  0.678(4) & 0.255(3) \\
        $N_c^\text{max}=196$ & 97.735(42) & 1.699(32) & 0.663(4) & 0.258(2) \\
         \hline
        \emph{ibmq\_sydney} &  &  &  &    \\
        qubits 8 \& 11 & 97.631(46) & 1.777(34) & 0.586(5) & 0.248(2) \\
        restless & 97.602(53) & 1.799(40) & 0.585(4) & 0.251(2) \\
        $N_c^\text{max}=196$ & 97.647(48) & 1.765(36) & 0.573(5) & 0.256(2) \\
         \hline
         \emph{ibmq\_montreal} &  &  &  &    \\
        qubit 0 & 99.931(1) & 0.035(1) & 0.486(3) & 0.499(3) \\
        standard & 99.941(2) & 0.029(1) & 0.494(6) & 0.491(6) \\
        $N_c^\text{max}=5101$ & 99.941(2) & 0.030(1) & 0.493(6) & 0.492(6)  \\
         \hline
         \emph{ibmq\_montreal} &  &  &  &    \\
        qubit 0 & 99.932(2) & 0.034(1) & 0.464(5) & 0.497(5) \\
        restless & 99.932(3) & 0.034(1) & 0.461(6) & 0.506(6)  \\
        $N_c^\text{max}=5101$ & 99.935(2) & 0.033(1) & 0.465(5) & 0.501(5) \\
         \hline
        \emph{ibmq\_montreal} &  &  &  &    \\
        qubits 1 \& 2 & 98.021(36) & 1.484(27) & 0.703(4) & 0.231(3) \\
        standard & 97.994(36) & 1.505(27) & 0.695(4) & 0.234(3)  \\
        $N_c^\text{max}=196$ & 97.976(37) & 1.518(28) & 0.693(5) & 0.236(3)  \\
         \hline
        \emph{ibmq\_montreal} &  &  &  &    \\
        qubits 1 \& 2 & 97.996(45) & 1.503(34) & 0.621(7) & 0.249(3) \\
        restless & 97.987(46) & 1.510(34) &  0.614(8) & 0.252(2)  \\
        $N_c^\text{max}=196$ & 98.038(51) & 1.472(38) &  0.619(7) &  0.247(3) \\
        \hline\hline
    \end{tabular}
\end{table}

\section{Device properties\label{sec:appendix_qubits}}

We ran the RB, calibration and QPT experiments on different qubits of various quantum devices based on system availability. This also shows that restless measurements are reproducible across different backends. Since restless measurements are sensitive to finite $T_1$-times we list the $T_1$-times of all qubits in Tab.~\ref{tab:devices}. For completeness, we also include the $T_2$-times, as well as the readout errors, that were reported by the backend at the time of the respective experiment. 

\begin{table}[htbp!]
    \centering
    \caption{Summary of qubit decay, coherence times and readout errors as reported by the backends.}
    \label{tab:devices}
    \begin{tabular}{l c c c} \hline\hline
       Device & $T_1$-time & $T_2$-time & read. error\\
      & $(\mu\mathrm{s})$ & $(\mu\mathrm{s})$ & (\%)  \\ \hline\hline
    Single-qubit RB & & & \\ \hline
     \emph{ibmq\_sydney} q13 & 134.5 & 156.3 & 1.70 \\
     \emph{ibmq\_montreal} q0 & 104.7 & 30.4 & 0.99 \\[0.5em] 
      Two-qubit RB & & & \\ \hline
      \emph{ibmq\_sydney} & & &  \\
      q8 & 105.4 & 173.8 & 3.74  \\
     q11 & 135.9 & 101.3 & 1.53   \\
     \emph{ibmq\_montreal} & & &  \\
     q1 & 102.9 & 23.2 & 1.66 \\
     q2 & 91.5 & 121.0 & 1.17 \\[0.5em] 
     Single-qubit cal. & & \\ \hline
     \emph{ibmq\_quito} q1 & 58.3 & 131.5 & 1.88  \\ 
     \emph{ibmq\_jakarta} q1 & 116.9 & 21.9 & 1.96  \\
     \emph{ibmq\_montreal} q3 & 83.0 & 70.7 & 0.74   \\[0.5em] 
     Two-qubit cal. & & \\ \hline
     \emph{ibmq\_jakarta} & & & \\
    q1 & 131.1 & 25.5 & 2.06   \\
     q3 & 117.7 & 36.3 & 2.48  \\[0.5em] 
     $\sqrt{X}$ calibration & & & \\ \hline
     \emph{ibmq\_bogota} q2 & 105.2 & 176.2 & 1.91  \\[0.5em] 
     Single-qubit QPT & & & \\ \hline
      \emph{ibmq\_sydney} & & & \\
    q3 & 106.5 & 98.8 & 1.87   \\
     q5 & 105.7 & 70.8 &  0.69  \\
     \emph{ibmq\_montreal} & & & \\
    q13 & 89.3 & 76.3 & 0.98   \\
     q14 & 91.8 & 74.2 & 1.58   \\
     \emph{ibmq\_manila} & & & \\
    q2 & 170.5 & 19.5 & 4.88   \\
     q3 & 172.9 & 65.8 & 1.69  \\[0.5em] 
     Two-qubit QPT & & & \\ \hline
     \emph{ibmq\_sydney} & & & \\
     q3 & 151.0 & 100.3 & 1.93   \\
     q5 & 122.4 & 90.4 &  2.42  \\
     \emph{ibmq\_montreal} & & & \\
    q13 & 56.8 & 30.6 & 2.22  \\
     q14 & 94.3 & 116.4 & 1.50   \\
     \emph{ibmq\_manila} & & & \\
    q2 & 160.9 & 19.0 & 4.0 \\
     q3 & 178.7 & 54.3 & 2.6  \\
     \hline\hline
    \end{tabular}
\end{table}

\section{Process tomography\label{sec:appendix_qpt}}

Here, we elaborate on the restless data processing for QPT.
When the qubits are initialized to $\ket{0}$ the input density matrices are always $\rho^\text{in}_j=U_j^\text{pre}\ket{0}\!\!\bra{0}\left(U_j^\text{pre}\right)^\dagger$.
In an ideal restless measurement on $n$ qubits the input density matrix is $\rho^\text{in}_{j,x}=U_j^\text{pre}\ket{x}\!\!\bra{x}\left(U_j^\text{pre}\right)^\dagger$ where $x\in\{0,1\}^n$.
However, in practice the measurement and the delay between the measurement and the next $U_j^\text{pre}$ induce errors which we model by an amplitude damping channel $\mathcal{E}_\text{ad}$.
In the single-qubit case the input states are given by equations~(\ref{eq:restless_prep_basis_0}) and~(\ref{eq:restless_prep_basis_1}) of the main text which trivially generalize to $n$ qubits.
This results in a total of $4^{n} \times2^n$ possible input states when only $4^{n}$ preparation rotations $U_j^\text{pre}$ are done.
Taking into account the three measurement basis there is a total of $4^{n} \times2^n\times3^{n}$ combinations of input states and measurement basis.
However, restless QPT only executes $4^{n}\times3^{n}$ circuits as does standard QPT.
The restless QPT data processor must therefore remap each measured shot to one of the $4^{n} \times2^n\times3^{n}$ combinations.
To post process restless QPT data we require the memory of each circuit as discussed in Sec.~\ref{sec:data_processing} of the main text.
Once again, we time order all the measurement outcomes.
If for $(U_j^\text{pre},U_i^\text{post})$ the previous measurement outcome was $\ket{0}$ we attribute the corresponding measurement to the input state $U_j^\text{pre}\ket{0}\!\bra{0}(U_j^\text{pre})^\dagger$ and post rotation $U_i^\text{post}$.
However, if the previous measurement outcome was $\ket{1}$ we attribute the measurement to the input state $U_j^\text{pre}[p\ket{0}\!\bra{0}-(1-p)\ket{1}\!\bra{1}](U_j^\text{pre})^\dagger$ and post rotation $U_i^\text{post}$.
This reasoning extends trivially to the multi-qubit case.
Since we now know the input state and measurement basis of each shot we can build up count dictionaries for each of the $4^{n} \times2^n\times3^{n}$ possible combinations.
The remainder of the analysis follows the standard QPT work flow.

\subsection{Additional process tomography data}

\begin{figure}[htbp!]
    \centering
    \includegraphics[width=1\columnwidth]{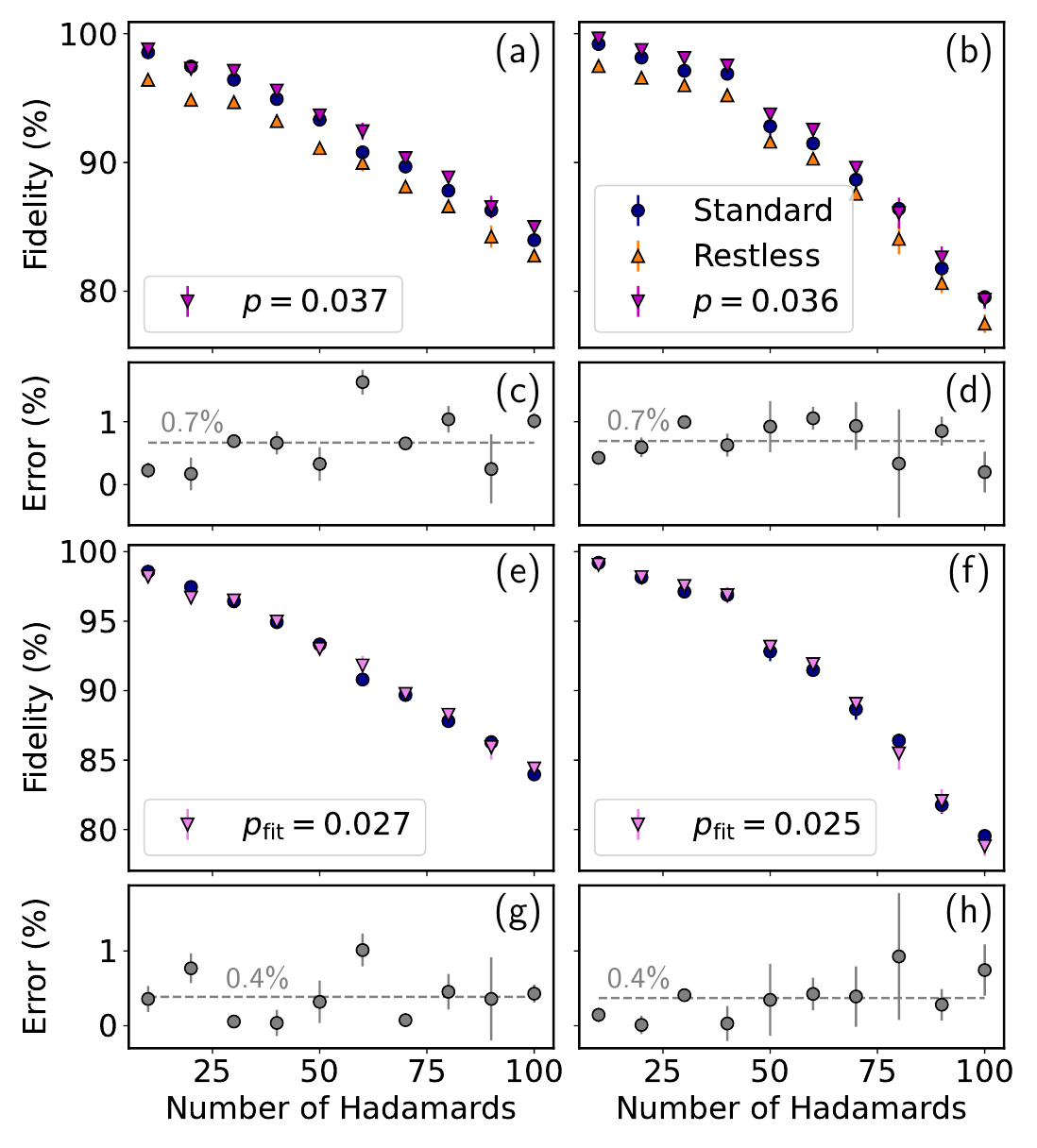}
    \caption{
    Single-qubit QPT on qubits 2 (a, c, e, g) and 3 (b, d, f, h) of \emph{ibmq\_manila}.
    (c) and (d) show the difference between the standard QPT data and the restless QPT data with $p$ obtained from $T_1$ times reported by the backend of $170.5~\mu{\rm s}$ and $172.9~\mu{\rm s}$, for qubits 2 and 3, respectively.
    Panels (e) and (f) show the best restless measured fidelity when treating $p$ as a fit parameter and panels (g) and (h) show the deviation $|\mathcal{F}_R(p_\text{fit})-\mathcal{F}_S|$.
    }
    \label{fig:restless_qpt_hadamard_manila}
\end{figure}

\begin{figure}[htbp!]
    \centering
    \includegraphics[width=1\columnwidth]{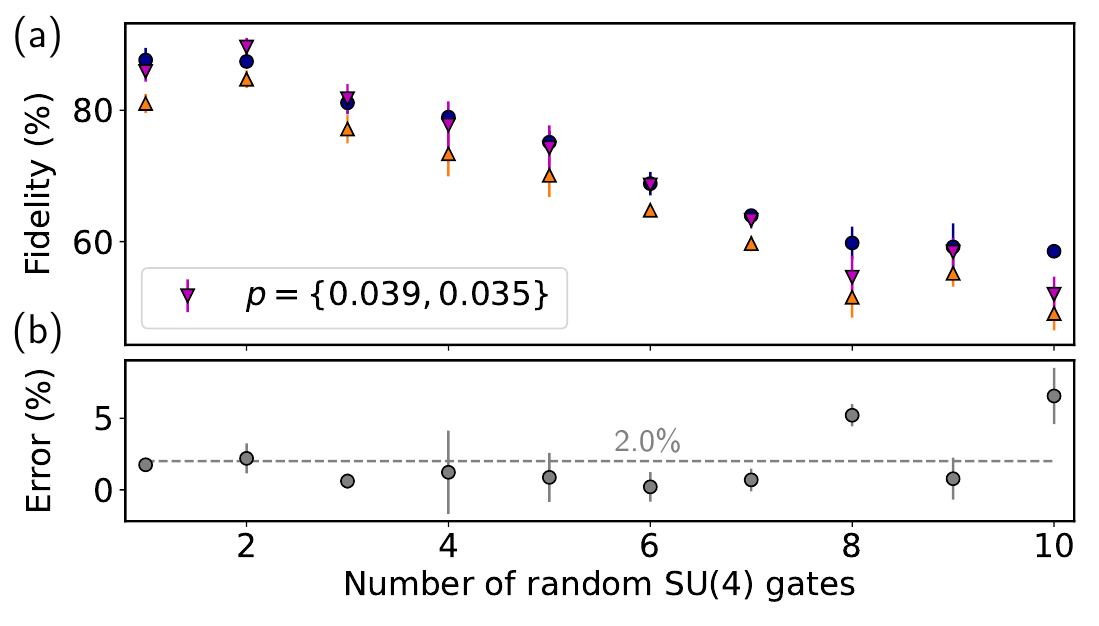}
    \caption{
    (a) QPT on random SU(4) gates between qubits 2 and 3 of \emph{ibmq\_manila}.
    The purple-down triangles show the restless fidelity with $p$ based on $T_1$-times reported by the backend of $160.9~\mu{\rm s}$ and $178.7~\mu{\rm s}$, for qubits 2 and 3, respectively.
    (b) Difference $|\mathcal{F}_R(p)-\mathcal{F}_S|$ between the fidelities measured with restless and active reset.
    }
    \label{fig:restless_qpt_cnots_manila}
\end{figure}

\begin{figure}[htbp!]
    \centering
    \includegraphics[width=1\columnwidth]{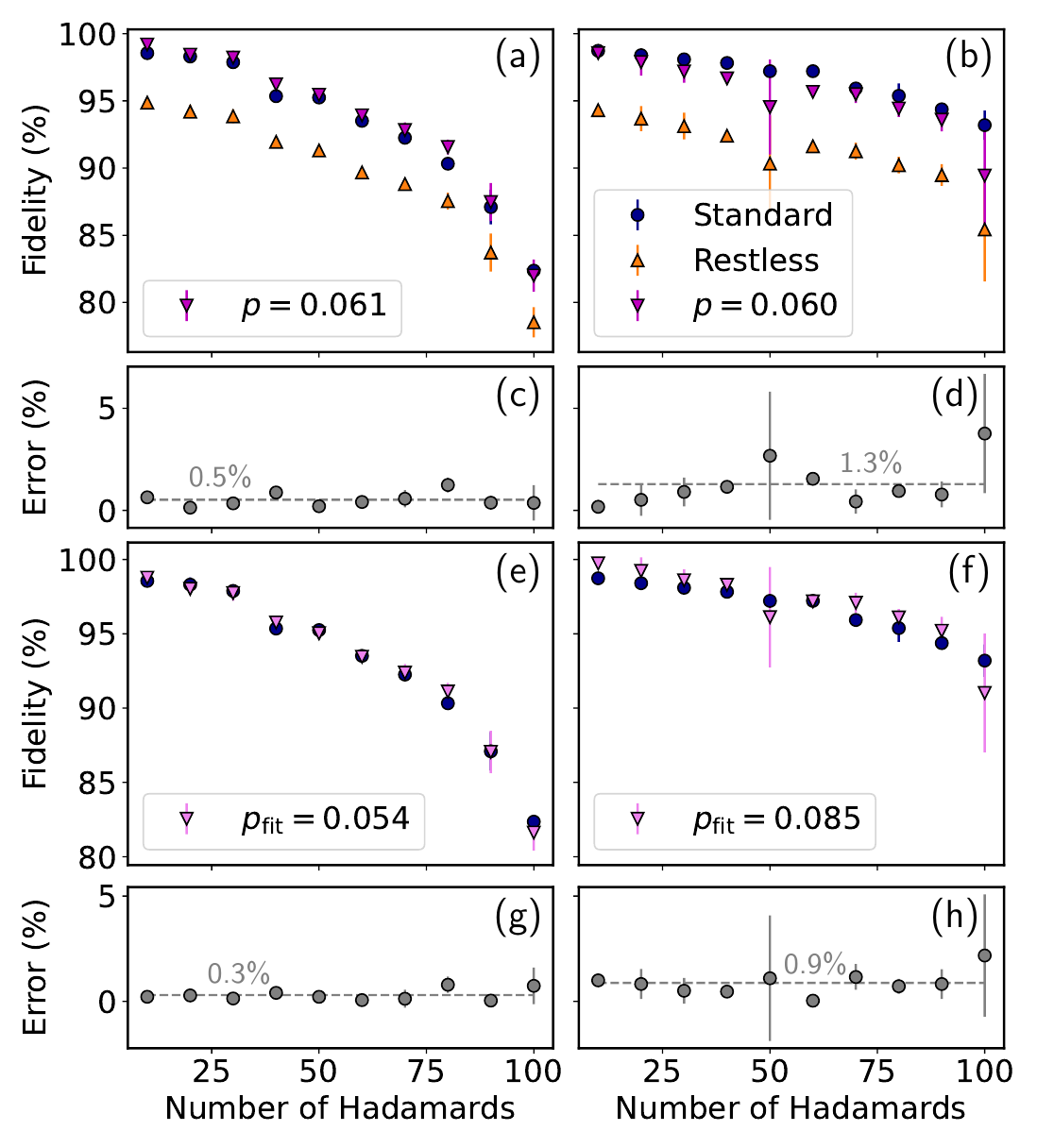}
    \caption{
      Single-qubit QPT on qubits 13 (a, c, e, g) and 14 (b, d, f, h) of \emph{ibmq\_montreal}.
      (c) and (d) show the difference between the standard QPT data and the restless QPT data with $p$ obtained from $T_1$ times reported by the backend of $89.3~\mu{\rm s}$ and $91.8~\mu{\rm s}$, for qubits 13 and 14, respectively.
      Panels (e) and (f) show the best restless measured fidelity when treating $p$ as a fit parameter and panels (g) and (h) show the deviation $|\mathcal{F}_R(p_\text{fit})-\mathcal{F}_S|$.
    }
    \label{fig:restless_qpt_hadamard_montreal}
\end{figure}

\begin{figure}[htbp!]
    \centering
    \includegraphics[width=1\columnwidth]{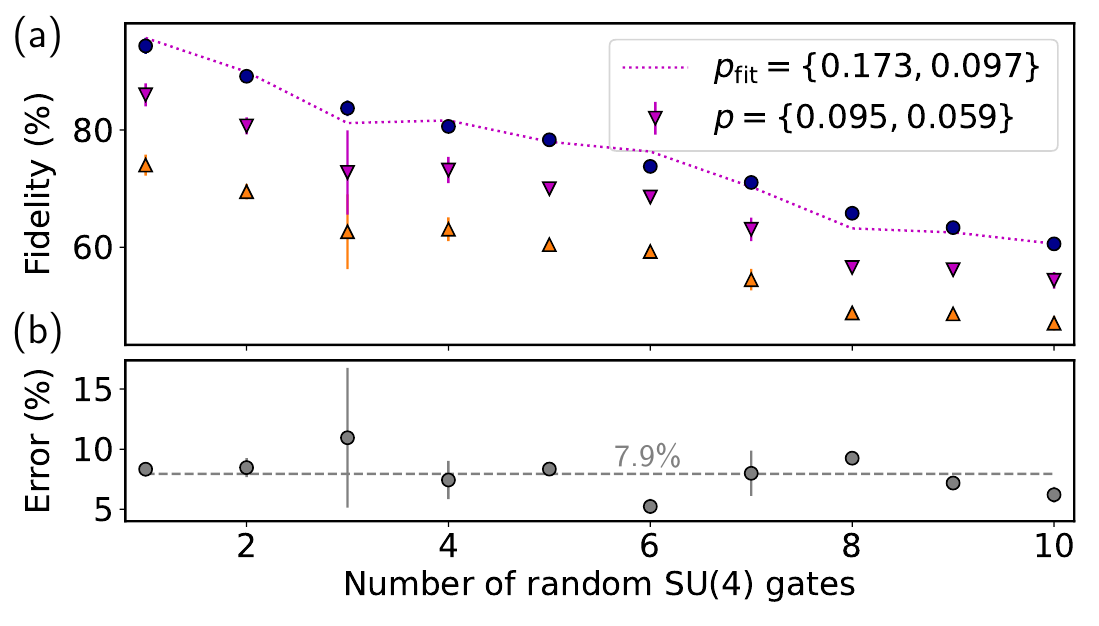}
    \caption{
    (a) QPT on random SU(4) gates between qubits 13 and 14 of \emph{ibmq\_montreal}.
    The purple-down triangles show the restless fidelity with $p$ based on $T_1$-times reported by the backend of $56.8~\mu{\rm s}$ and $94.3~\mu{\rm s}$, for qubits 13 and 14, respectively.
    The dotted purple line indicates the restless measured fidelities when treating the decay probabilities $p$ as fit parameters.
    (b) Difference $|\mathcal{F}_R(p)-\mathcal{F}_S|$ between the fidelities measured with restless and active reset.
    The deviation of on average $7.9 \pm 1.3\%$ is reduced to $\langle |\mathcal{F}_R(p_\text{fit})-\mathcal{F}_S| \rangle = 1.3 \pm 1.3 \%$ for $p_\text{fit} = \{0.173, 0.097\}$.
    }
    \label{fig:restless_qpt_cnots_montreal}
\end{figure}

In addition to the data presented in the main text we also run standard and restless QPT on \emph{ibmq\_manila}, see Fig.~\ref{fig:restless_qpt_hadamard_manila} and \ref{fig:restless_qpt_cnots_manila}, and \emph{ibmq\_montreal}, see Fig.~\ref{fig:restless_qpt_hadamard_montreal} and \ref{fig:restless_qpt_cnots_montreal}.
We characterized an even sequence of Hadamard gates, from ten to 100, and a sequence of random CNOT decomposed SU(4) gates from one to ten.
We run each QPT measurement with 4096 shots and each point is the average of three individual measurements with the standard deviation shown as error bars.
Readout error mitigation is used. 
In Fig.~\ref{fig:restless_qpt_hadamard_manila} to \ref{fig:restless_qpt_cnots_montreal} the blue circles indicate standard QPT measurements, the orange up-triangles indicate restless QPT measurements by reassigning the shots to the basis $\{Z_m, Z_p, X_m, X_p, Y_m, Y_p\}$, and the purple down-triangles show the same data but processed with the restless preparation basis corresponding to Eq.~(\ref{eq:restless_prep_basis_0}) and (\ref{eq:restless_prep_basis_1}) in the main text.

As in the main text, we observe that the restless QPT data analyzed under the assumption that restless measurements are ideal, i.e. when the previous outcome was $\ket{1}$ the initial state is $\ket{1}$, underestimate the gate fidelity.
Furthermore, we observe that a finite $p$ obtained from $T_1$, measurement, and delay times mitigates state preparation errors when using Eq.~(\ref{eq:restless_prep_basis_0}) and (\ref{eq:restless_prep_basis_1}) of the main text as input states.
Furthermore, a few restless QPT measurements are biased by large outliers.
The exact source of these outliers is unknown but could be due to measurement or $T_1$ related variations.

\clearpage

\bibliography{references}

\end{document}